\documentclass[fleqn,usenatbib]{rasti}

\usepackage{newtxtext,newtxmath}

\usepackage[T1]{fontenc}

\DeclareRobustCommand{\VAN}[3]{#2}
\let\VANthebibliography\thebibliography
\def\thebibliography{\DeclareRobustCommand{\VAN}[3]{##3}\VANthebibliography}

\usepackage{graphicx}	
\usepackage{amsmath}	
\usepackage[version=4]{mhchem}
\usepackage{pdflscape}
\usepackage{float}

\title[Astroclimes]{\textit{Astroclimes} - measuring the abundance of \ce{CO2} and \ce{CH4} in the Earth's atmosphere using astronomical observations}

\author[M. A. F. Keniger et al.]{
Marcelo Aron F. Keniger,$^{1,2}$\thanks{E-mail: Marcelo-Aron.Fetzner-Keniger@warwick.ac.uk}
David Armstrong,$^{1,2}$
Matteo Brogi,$^{3,4}$
Siddharth Gandhi$^{1,2}$ and 
\newauthor
\hspace{0.1cm}Marina Lafarga$^{1,2}$
\\
$^{1}$Department of Physics, University of Warwick, Gibbet Hill Road, Coventry CV4 7AL, UK \\
$^{2}$Centre for Exoplanets and Habitability, University of Warwick, Gibbet Hill Road, Coventry CV4 7AL, UK \\
$^{3}$Dipartimento di Fisica, Università degli Studi di Torino, via Pietro Giuria 1, I-10125, Torino, Italy \\
$^{4}$INAF-Osservatorio Astrofisico di Torino, Via Osservatorio 20, I-10025 Pino Torinese, Italy \\
}

\date{Accepted XXX. Received YYY; in original form ZZZ}

\pubyear{\the\year{}}

\begin{document}
\label{firstpage}
\pagerange{\pageref{firstpage}--\pageref{lastpage}}
\maketitle

\begin{abstract}
Monitoring the abundance of greenhouse gases is necessary to quantify their impact on global warming and climate change. Carbon dioxide (\ce{CO2}) and methane (\ce{CH4}) are the two most important greenhouse gases when it comes to global warming. Although a number of satellites and ground-based networks measure the total column volume mixing ratio (VMR) of these gases, they rely on sunlight, and column measurements at night are comparatively scarce. We present a new algorithm, \textit{Astroclimes}, that hopes to complement and extend nighttime \ce{CO2} and \ce{CH4} column measurements. \textit{Astroclimes} can measure the abundance of greenhouse gases on Earth by generating a model telluric transmission spectra and fitting it to the spectra of telluric standard stars in the near-infrared taken by ground-based telescopes. An extensive dataset from the CARMENES spectrograph in the Calar Alto Observatory was compiled, which included all of the publicly available data from 2016 to 2024, as well as new observations carried out alongside a weather balloon launch. A Markov Chain Monte Carlo (MCMC) analysis on this extensive dataset showed that \textit{Astroclimes} was able to recover the long term trend known to be present in the molecular abundances of both \ce{CO2} and \ce{CH4}, but not their seasonal cycles. Using the Copernicus Atmosphere Monitoring Service (CAMS) global greenhouse gas reanalysis model (EGG4) as a benchmark, we identified an overall vertical shift in our data and quantified the long-term scatter in our retrievals. We found that our ground level and column-averaged \ce{CO2} dry mole fractions (DMFs) exhibit a scatter of $\pm 5$ ppm and $\pm 9$ ppm, respectively. For \ce{CH4}, these values were $\pm 31$ ppb and $\pm 42$ ppb, respectively. The vertical shift was 14 ppm and 15 ppm for the \ce{CO2} ground level and column-averaged DMFs, respectively, and 42 ppb and 7 ppb for the \ce{CH4} ground level and column-averaged DMFs, respectively. The scatter on a 1 hour timescale, however, is much lower, of order $\pm 1$ ppm and $\pm 10$ ppb for ground \ce{CO2} and \ce{CH4}, respectively, which is on par with the uncertainties on individual measurements. Although currently the precision of the method is not in line with state of the art techniques using dedicated instrumentation, which achieve sub-ppm precision, it shows promise for further development. 
\end{abstract}




\section{Introduction}\label{introduction}

Climate change is one of the most pressing matters in today's society, having already caused substantial damage to ecosystems and leaving a significant portion of the population highly vulnerable to climatic hazards, according to the latest report from the Intergovernmental Panel on Climate Change \citep[IPCC;][]{lee_IPCC2023}. Even though the vast majority of climate scientists agrees that humans are responsible for the increase in global temperature that drives climate change \citep{cook16}, current policies are still not enough to decrease the emission of greenhouse gases (GHG), which are the cause of this rise in temperature \citep[e.g. ][]{peters2019,lee_IPCC2023}. Among the GHGs, the ones that dominate global warming are carbon dioxide (\ce{CO2}) and methane \citep[\ce{CH4};][]{lee_IPCC2023}. As such, obtaining accurate and reliable measurements for the abundance of said gases in the atmosphere is paramount to understanding and modelling climate change, as well as to guide governmental policies for alleviating its impact \citep{bruhwiler2021}.

Ground-based carbon dioxide measurements are usually taken using a Non-Dispersive Infrared (NDIR) analyser (see \citealt{komhyr1989} or \citealt{hodgkinson2013a,hodgkinson2013b,jha2022} for reviews), but another technique called Cavity Ring-Down Spectroscopy \citep[CRDS;][]{wheeler1998,berden2000} has also been employed. The basic principle behind the two methods is similar and consists in analysing the transmittance of an infrared (IR) radiation source, such as an LED \citep{jha2022} or a laser \citep{wheeler1998}, as it goes through a chamber with a sample gas. Each gas molecule will affect the transmittance in a different way, so they can be distinctively identified by analysing the absorption characteristics \citep{jha2022}. In the NDIR method, the magnitude of absorption is used to infer the abundance of gas present, whereas the CRDS method is based on the rate of absorption instead \citep{berden2000}.

The longest record of direct measurements of \ce{CO2} in the atmosphere comes from the Mauna Loa Observatory in Hawaii, with ongoing observations since the 1950s \citep[e.g. ][]{pales1965,keeling1976,bacastow1985,komhyr1989,thoning1989}.  At first, these measurements were carried out solely by the Scripps Institution of Oceanography \citep[SIO;][]{pales1965}, but since 1974 the Global Monitoring Laboratory (GML)\footnote{\href{https://gml.noaa.gov/}{https://gml.noaa.gov/}}, previously known as the Geophysical Monitoring for Climate Change (GMCC) program, has also been carrying out continuous \ce{CO2} measurements in Mauna Loa \citep{komhyr1989,thoning1989}. The GML is a global network that monitors atmospheric parameters such as the abundance of \ce{CO2} and other GHGs. Besides Mauna Loa, there are three other baseline observatories (American Samoa, Alaska and Antarctica), in addition to measurements from tall towers \citep{bakwin1998,peters2007,andrews2014}, small aircraft \citep{sweeney2015}, weather balloons \citep{karion2010} and more than 50 sites spread worldwide \citep{conway1994,ballantyne2012}\footnote{\href{https://gml.noaa.gov/ccgg/about.html}{https://gml.noaa.gov/ccgg/about.html}}. Initially, the GML employed the NDIR technique to measure \ce{CO2} abundances in Mauna Loa, but since 2019 they switched to a detector that uses the CRDS technique\footnote{\href{https://gml.noaa.gov/ccgg/about/co2_measurements.html}{https://gml.noaa.gov/ccgg/about/co2\textunderscore measurements.html}}. On their website, one can find reported uncertainties for the measurements of \ce{CO2} and \ce{CH4}, which are stated to be between 0.1-0.2 ppm and 1-3 ppb, respectively\footnote{\href{https://gml.noaa.gov/dv/iadv/help/ccgg_details.html}{https://gml.noaa.gov/dv/iadv/help/ccgg\textunderscore details.html}}.

Since the start of the continuous observations of atmospheric \ce{CO2} in Mauna Loa, certain trends have become evident: a seasonal variation and a long term increase \citep[e.g. ][]{pales1965,keeling1976,komhyr1989,thoning1989}. The seasonal variation is caused by the uptake and release of \ce{CO2} by the land biosphere \citep{junge1968}. The long term increase has been attributed to anthropogenic global emissions of \ce{CO2} \citep[e.g. ][]{watts1980,keeling1985,thoning1989} which are dominated by the combustion of fossil fuels \citep[e.g. ][]{peters2019,lee_IPCC2023}. This increase in \ce{CO2} in the atmosphere is a direct cause of the observed rise in global temperature \citep[e.g. ][]{hansen2010,voosen2021}, which had been predicted by a number of climate models \citep[e.g ][]{watts1980,cox2000,friedlingstein2006}. 

On top of the seasonal and long term trend, the abundance of \ce{CO2} in the atmosphere can also be affected by climate effects such as the El Niño-Southern Oscillation \citep[ENSO; e.g. ][]{bacastow1980,mcphaden2006}. In El Niño years (the warm phase of ENSO), global \ce{CO2} concentrations typically increase, as droughts and high temperatures contribute to more fires and to the change in balance between plant respiration and photosynthesis, reducing the carbon uptake by land vegetation \citep{mcphaden2006}. For reference, \cite{betts2016} forecast that the \ce{CO2} growth rate between 2015 and 2016, which experienced a strong El Niño, would be 3.15~ppm~yr$^{-1}$, almost 1~ppm larger than the previous year. \cite{chatterjee2017} and \cite{patra2017} confirmed this forecast. 2023, also an El Niño year, had a \ce{CO2} growth rate of $2.79\pm0.08$~ppm, almost 30\% larger than the previous year growth rate and the fourth largest since 1959, falling behind 2015, 2016 and 1998, all of which were strong El Niño years \citep{friedlingstein2025}.

To keep on monitoring the state of the climate and to assess the reliability of climate models, it is crucial that continuous atmospheric \ce{CO2} measurements are maintained. The \ce{CO2} measurements from the four GML Baseline Observatories have the advantage of being taken in remote locations in different parts of the globe, so they accurately represent the background atmosphere and serve as the backbone of the GML's climate monitoring \citep{stanitskiGMDresearchplan}. However, since these measurements are taken at ground level, they have limited information regarding the total column abundance of \ce{CO2} in the atmosphere, which adds uncertainty to the study of the growth rate of atmospheric \ce{CO2} \citep{friedlingstein2023}. The aforementioned aircraft measurements have the capability of doing so, but they only go up to $\sim 13$km and are taken mostly inside the United States\footnote{\href{https://gml.noaa.gov/ccgg/aircraft/}{https://gml.noaa.gov/ccgg/aircraft/}}.

For column measurements of \ce{CO2}, satellites such as NASA's Orbiting Carbon Observatory 2 \citep[OCO-2;][]{basilio14, crisp17} and, more recently, OCO-3 \citep{eldering19} were launched, as well as the Greenhouse gases Observing SATellite 2 (GOSAT-2), by the Japanese Aerospace Exploration Agency \citep[JAXA;][]{imasu2023}, which also includes column measurements of \ce{CH4}. These satellites observe sunlight reflected on the Earth's surface and measure the \ce{CO2} abundance by analysing the solar spectra in the near-infrared (NIR). 

There are also ground-based networks focused on measuring the column abundances of relevant greenhouse gases such as \ce{CO2} and \ce{CH4}, one of them being the Total Carbon Column Observing Network \citep[TCCON;][]{wunch2011}, whose measurements are directly comparable to those from space-based instruments and thus can provide a link between satellite and ground-based measurements \citep{wunch2011}. The latest reported uncertainties for TCCON measurements of the column-averaged \ce{CO2} and \ce{CH4} are 0.16\% ($\sim0.6$ ppm) and 0.4\% ($\sim 7$ ppb), respectively, for solar zenith angles less than $82^\circ$ \citep{laughner23b}. When compared to collocated TCCON measurements, column-averaged \ce{CO2} estimates from OCO-2 and OCO-3 show root mean squared errors (RMSEs) of around 0.8 and 0.9 ppm, respectively \citep{taylor2023}.


More recently, nighttime measurements of the column abundance of \ce{CO2} started being carried out by the Atmospheric Environment Monitoring Satellite \citep[AEMS;][]{pei2023} and NASA's Active Sensing of \ce{CO2} Emissions over Nights, Days, and Seasons \citep[ASCENDS;][]{mao2024}. Both missions employ an integrated-path differential absorption (IPDA) Light Detection and Ranging (LiDAR) system, which in addition to allowing column measurements at night, also works on cloudy conditions and high latitudes, which the previous networks struggle with.

An alternative to measurement networks comes in the form of reanalysis models. These combine past weather measurements with current climate models to create a global and complete view of our atmosphere that fills the gaps on the measurement networks. The latest reanalysis model available for greenhouse gases is provided by the Copernicus Atmosphere Monitoring Service \citep[CAMS;][]{inness2019} and is called EGG4 \citep{agusti-panareda2023}. As of March 2025, the CAMS EGG4 reanalysis model is available from 2003 to 2020. 

Here, we describe a novel method for measuring the column abundances of \ce{CO2} and \ce{CH4} that aims to complement and extend nighttime measurements. Our measurement approach is similar to that of OCO-2 and TCCON, but instead of using sunlight, we analyse the spectra of stars taken by ground-based telescopes. The selected stars for our analysis are known as telluric standard stars \citep{vacca2003,seifahrt2010}, usually O, B or A type stars that are called telluric standards because they are very hot, thus have very few spectral lines, and rotate very fast, thus the few lines that they do have are broadened, making the stellar lines easy to distinguish from the narrower telluric lines \citep{ulmermoll19}.

Telluric standards are a common by-product of ground-based astronomical spectroscopic observations. High resolution spectroscopy is used in astronomy, for example, to detect exoplanets through the radial velocity (RV) method \citep[e.g. ][]{mayor1995,lovis2006} and to study exoplanet atmospheres \citep[e.g. ][]{brogi2012,brogi2013,brogi2014,madhusudhan2014,crossfield2015}. Removal of the telluric lines is essential in both cases, but more so for exoplanet atmosphere studies, since the telluric signal can be orders of magnitude higher than the planet signal \citep{brogi2014}. So that the telluric lines can be better removed, and in our case, so that they can be properly modelled, it is important that the observational spectra are of high enough resolution such that individual features in the telluric lines can be resolved.

Historically, telluric lines have been removed from stellar spectra using the standard star method \citep[e.g. ][]{vidalmadjar1986,vacca2003}, which consists of dividing the science spectra by the spectra of a telluric standard star. This technique, however, requires the standard star to be observed close in time and airmass to the science target \citep{vacca2003}, so that atmospheric conditions are as similar as possible, thus costing telescope time that could otherwise be used for other scientific purposes. Additionally, no matter how void of spectral lines a star's spectra may be, it is never featureless, which will inevitably affect the resulting science spectra \citep{lallement93,bailey2007,ulmermoll19}.

An alternative telluric removal method has been put forward by many authors \citep[e.g. ][]{lallement93,bailey2007,seifahrt2010,cotton2014} and is known as the synthetic transmission method. As the name suggests, instead of using the spectra of a telluric standard, a model telluric spectra is computed, and the telluric lines are removed by dividing the science spectra by this model telluric spectra. Generating a synthetic telluric spectra usually requires atmospheric parameters such as pressure, temperature and molecular abundances as a function of height, coupled with a molecular line database and a radiative transfer model \citep{seifahrt2010,ulmermoll19}.

There are many tools in the literature made for removing telluric lines using the synthetic transmission method, such as \textit{Molecfit} \citep{smette15, kausch2015}, \textit{TelFit} \citep{gullikson2014} and \textit{TAPAS} \citep{bertaux2014}. These three telluric correction codes are analysed and compared to each other and also to the standard star method in \citet{ulmermoll19}, where they conclude that \textit{Molecfit} is the most complete package between the three, and that synthetic transmission has some advantages over the standard star method when dealing with water lines, but performs worse for oxygen lines.

All of these tools, however, were made primarily to remove the telluric lines, not to study them, so often their fitting approach is a simple Levenberg-Marquardt least-squares fit \citep{gullikson2014,smette15} and uncertainties for the molecular abundances are not reported. Thus, we created our own synthetic transmission code, called ``\textit{Astroclimes}''. Developing our own code also grants us more control over the modelling process and allows us to tailor its performance to our specific needs, which in this case is to measure the abundance of greenhouse gases in the Earth's atmosphere, setting it apart from the other telluric removal tools in the literature. The capabilities of our code for removing telluric lines will be tackled in future work. 

By utilising telluric standard stars as opposed to the Sun, we can perform column measurements at night, which could provide insights on the natural cycles of the studied molecules. Additionally, since telluric standards are a common by-product of ground-based astronomical spectroscopic observations, there are archival data at our disposal in telescopes spread across the world, thus providing a potential new measurement network that could be used to complement current global climate models.

This paper is structured as follows: first, we explain the step-by-step of our new method in Section \ref{sec_methodology}; then, in Section \ref{sec_data}, we describe all of the different data sets employed in our analysis and how they were handled; Section \ref{sec_results} contains our results and discussion, where we explore the influence of using different atmospheric profiles and test our model against the Cerro Paranal Advanced Sky Model \citep[ASM;][]{noll2012,jones2013}, and against the CAMS global greenhouse gas reanalysis model; and finally in section \ref{conclusions} we present our conclusions. 

\section{Methodology}\label{sec_methodology}
\subsection{Generating the model spectra}
The first step of the algorithm presented here is to generate a model telluric spectrum that contains absorption lines of certain molecular species present in the Earth's atmosphere. The two main ingredients needed to compute this model spectrum are the cross-sections of the desired molecules and an atmospheric profile. 

The cross-sections were calculated in a similar fashion as in \cite{gandhi17} and \cite{gandhi20}. However, in these papers they calculated a grid of molecular cross-sections for volatile species found in giant planet atmospheres in a ``high temperature'' regime, which goes from 300-3500 K. The Earth's atmosphere is below this regime, so a new grid was calculated specifically for this work for a ``low temperature'' regime. This grid contains cross-section values for 11 different pressures ranging from 0-5 log(Pa) in 0.5 log(Pa) increments, 11 different temperatures ranging from 100-350 K in 25 K increments, and 2480001 wavelength values ranging from 0.4-50 $\mu$m in constant wavenumber steps of 0.01 cm$^{-1}$. The molecules available are \ce{CO2}, \ce{CH4}, \ce{H2O}, \ce{O2}, \ce{N2}, \ce{CH3Cl}, \ce{CO}, \ce{H2}, \ce{HCN}, \ce{N2O}, \ce{NO2}, \ce{O3} and \ce{OH}. To compute all of the cross-sections, the line lists from the HITRAN database \citep{gordon2020} were used.

The atmospheric profile describes how the pressure $P$, the temperature $T$ and the molecular abundances $x_i$ for each molecule $i$ vary as a function of height. The process of obtaining the atmospheric profile is explained in Section \ref{sec_atmdata}. The atmospheric profile gives the aforementioned parameters for certain height levels. One atmospheric ``layer'' is the space between said levels. To quantify the amount of light that is absorbed as it passes through each layer, the opacity needs to be calculated, which for molecular line absorption is given by:

\begin{equation}\label{tau}
    \tau _z (\lambda) = nr \sum_{i} x_i \sigma _i (\lambda)
\end{equation}
where $\sigma _i (\lambda)$ is the cross-section of molecule $i$ as a function of wavelength $\lambda$, 
\begin{equation}\label{eq_idealgaslaw}
    n = \frac{P}{k_\text{B} T}
\end{equation}
is the number density of particles in the atmosphere, with $k_\text{B}~=~1.380649 \times 10^{-23}$~m$^2$~kg~s$^{-2}$~K$^{-1}$ being the Boltzmann constant, and 
\begin{equation}
    r = \frac{Z_\text{top} - Z_\text{bottom}}{\cos{\theta}}
\end{equation}
is the distance travelled by the light inside that layer, where $Z_\text{top}$ refers to the height at the top limit of the layer, $Z_\text{bottom}$ refers to the height at the bottom limit of the layer, and $\theta$ is the angle between the target and the local zenith. $\theta$ is related to the airmass by the following equation:
\begin{equation}
    \cos{\theta} = \frac{1}{\text{airmass}}
\end{equation}

To calculate the opacity $\tau _z (\lambda)$ representative of each atmospheric layer, certain values of the pressure, temperature, molecular abundance and cross-section must be chosen. They could be the values at the top of the layer limit, at the bottom of the layer limit, or somewhere in between. Currently, the representative values for $P$, $T$ and $x_i$ taken are the mid-point of each layer, which assumes a linear relation between them and the altitude, but there are plans to refine that in future work to take into account the non-linear relation between the variables, as can be seen from Figures \ref{fig_atmospheric_profiles} and \ref{fig_atmospheric_profiles_ggg2020}. From the pressure and the temperature, a value for the cross-section is linearly interpolated from the cross-sections grid, and thus the opacity can be calculated for a specific atmospheric layer, which is indicated by the index $z$ in Equation (\ref{tau}), starting at the top of the atmosphere $h_\text{top}$ and going all the way down to the observatory height $h_0$. The transmission $T(\lambda)$ is then calculated from the sum of the opacity in all layers by the Beer-Lambert law:

\begin{equation}\label{trans}
    T(\lambda) = e^{-\sum_{h_{\text{top}}}^{h_{0}} \tau _z (\lambda)}
\end{equation}

Apart from line absorption by molecular species in the atmosphere, there are other effects caused by the interaction of light with particles that can change the transmission spectra. Two such effects are described in \cite{noll2012}, where the authors describe the Cerro Paranal Advanced Sky Model, an atmospheric radiation model for Cerro Paranal, one of the European Southern Observatory's (ESO)\footnote{\href{https://www.eso.org/public/}{https://www.eso.org/public/}} astronomical observatories in Chile. These effects are Rayleigh scattering and aerosol scattering, which can be parametrised by the following expressions, respectively:

\begin{equation}\label{rayscat}
    \tau _R (\lambda) = \frac{P_0}{1013.25} \left ( 0.00864 + 6.5 \times 10^{-6} h_0 \right )
    \times \lambda ^{-(3.916 +0.074\lambda + \frac{0.050}{\lambda})}
\end{equation}

\begin{equation}\label{aerososcat}
    T_\text{aerosol} (\lambda) = 10^{-0.4k(\lambda) \times \text{airmass}} \quad ,
\end{equation}
where
\begin{equation}\label{aeroext}
    k(\lambda) = k_0 \lambda ^{\alpha} \quad ,
\end{equation}
where $k_0 = 0.013 \pm 0.002$ mag/airmass and $\alpha = -1.38 \pm 0.06$, with the wavelength $\lambda$ in $\mu$m. In Equation (\ref{rayscat}), $P_0$ is the site pressure in hPa and $h_0$ is the site height in km. It should be noted that Equation (\ref{rayscat}) corresponds to an equivalent opacity, whereas Equation (\ref{aerososcat}) is an equivalent transmission.

Another effect that is particularly important in the region containing the \ce{O2} lines is collision-induced absorption (CIA), described in \cite{gordon17} and \cite{karman19}. This type of absorption occurs when two species experience a close encounter and give rise to a transient or interaction-induced dipole moment \citep{hubeny2014}. Since this is a second order interaction, its opacity is calculated by:

\begin{equation}\label{taucia}
    \tau _\text{CIA} (\lambda) = (nr)^2 \sum_{i,j} x_{i}x_{j}  \alpha_{i,j} (\lambda) \quad ,
\end{equation}
where $\alpha_{i,j} (\lambda)$ are the coefficients associated with a certain molecular collision involving molecules $i$ and $j$. These coefficients can be obtained from the HITRAN database\footnote{\href{https://www.hitran.org/cia/}{https://www.hitran.org/cia/}} \citep{rothman2013} for a number of molecular collisions. For this work, the absorption caused by the collision of \ce{O2} molecules with other atmospheric molecules was included. The HITRAN database has a special file for this, named ``\ce{O2}-air'', which includes \ce{O2}-\ce{O2}, \ce{O2}-\ce{N2} and \ce{O2}-\ce{Ar} collisions. The coefficients in these files are measured for a certain wavelength range and a certain temperature. The coefficients to be used in the modelling are obtained via linear interpolation if the atmospheric temperature is inside the coefficient's temperature range, otherwise the coefficient of the closest temperature value is used.

The final transmission spectra, with all of the aforementioned effects included, is given by:

\begin{equation}\label{full_trans}
    T(\lambda) = e^{-\sum_{h_{\text{top}}}^{h_{0}} \tau _z (\lambda)} \times e^{-\sum_{h_{\text{top}}}^{h_{0}} \tau _\text{CIA} (\lambda)} \times e^{-\tau _R (\lambda)} \times T_\text{aerosol} (\lambda)
\end{equation}

The model given by Equation (\ref{full_trans}) has not yet accounted for instrumental broadening, which is a necessary step that has to be done before comparing the model to observations. This is done by the convolution of the unbroadened spectra and a kernel that describes the line spread function, that is, the shape of the spectral lines. Here, a Gaussian kernel is used to model the lines, and its width depends on the resolution of the instrument where the data comes from. An important step in this process is that the wavelength distribution must first be converted to have a constant resolving power $R = \frac{\lambda}{\Delta \lambda}$ before taking the convolution.

\subsection{Comparing model and data}\label{sec_comp_mod_data}
With that, the model is ready to be compared to observational spectroscopic data. To quantify the agreement between model and observation and fit for the molecular abundances, the approach described in \cite{brogiline} was employed. In their framework, they build a likelihood function $L$ starting from Pearson's cross-correlation coefficient via the following equation:

\begin{equation}\label{eq_loglike}
    \log{(L)} = - \frac{N}{2} \left [s_{f}^{2} - 2 R(s) + s_{g}^{2} \right ]
\end{equation}

In Equation (\ref{eq_loglike}), $N$ is the total number of data points, $s_{f}^{2}$ is the variance of the observed spectrum $f(n)$, $s_{g}^{2}$ is the variance of the model spectrum $g(n)$ and $R(s)$ is the cross-covariance, given by, respectively:

\begin{equation}\label{eq_sf2}
    s_{f}^{2} = \frac{1}{N} \sum_{n} f^{2}(n)
\end{equation}

\begin{equation}\label{eq_sg2}
    s_{g}^{2} = \frac{1}{N} \sum_{n} g^{2}(n-s)
\end{equation}

\begin{equation}\label{eq_Rs}
    R(s) = \frac{1}{N} \sum_{n} f(n)g(n-s)
\end{equation}

In the previous equations, $n$ refers to each wavelength value in the spectra and $s$ refers to a wavelength shift. In this case, there is no shift $s$ between the models and the data because the telluric lines are not subject to any significant Doppler shift with respect to the observer. Hence, we do not include the shift as a model parameter and instead fix $s = 0$. Equation (\ref{eq_loglike}) is then used to drive a Markov-Chain Monte Carlo (MCMC) to estimate the best-fit model parameters, with details about chain length and convergence criteria explained in Section \ref{sec_results}.

\section{Data}\label{sec_data}

\subsection{Astronomical spectroscopic observations}\label{sec_specobs}
The key molecules of interest, \ce{CO2} and \ce{CH4}, have spectral lines in the near-infrared, which ranges from 0.9-2.5 $\mu$m. There are several suitable instruments spread across the globe, for example CARMENES \citep{quirrenbach14} in the 3.5m telescope at the Calar Alto Observatory, Spain, CRIRES+ \citep{follert2014,dorn2023}, located at the VLT in Cerro Paranal, Chile, NIRPS \citep{bouchy2017,artigau2024} at the La Silla Observatory, also in Chile, iSHELL \citep{rayner2016,rayner2022} at the NASA InfraRed Telescope Facility (IRTF) in Mauna Kea, Hawaii, SPIRou \citep{artigau2014} at the Canada-France-Hawaii-Telescope, also in Mauna Kea, and GIANO \citep{olivia2006,origlia2014} at the Telescopio Nazionale Galileo (TNG) at the Roque de Los Muchachos Observatory in La Palma. All of these instruments operate in the high-resolution spectroscopic regime, with resolutions ranging from $R\sim 50000-100000$. 

While getting observation time in these telescopes can be competitive, the data required for this work has the advantage of being a common by-product of many observation proposals that employ spectroscopic data. The type of stars best suited for the method proposed here are telluric standard stars, which are typically observed for calibration purposes, making it easier to obtain data that may not be publicly available (usually, astronomical observatories contain a proprietary period before making the data public, which in the case of CARMENES is one year). Telluric standard stars are often very bright, so they allow for short exposure times, usually of order seconds to minutes. Therefore, not only is there a large sample of archival data available, but being granted time in these telescopes for new observations is a viable option. For this work, both of these pathways were explored, with the CARMENES spectrograph being the instrument of choice as a first test for \textit{Astroclimes}. 

CARMENES is mounted on the 3.5m telescope at the Calar Alto Observatory (CAHA), located in the southern part of Spain (coordinates $37.22^\circ$N and $2.55^\circ$W, altitude 2168 m). The choice of CARMENES as the instrument to first test \textit{Astroclimes} stems from the fact that its location was suitable for a weather balloon launch, described in Section \ref{sec_weather_balloon}, plus there was also the opportunity to perform new observations simultaneously to the weather balloon launch. These new observations were carried out on the night of April 24th 2023. 

As a target, a telluric standard star was chosen. Based on the visibility conditions at the time interval granted to us, the selected target was HR 5676, an A type star with $V = 5.272$ \citep{hog2000}. In total, 73 observations were carried out throughout the night, between UT 20h36 and UT 03h57, with 68 seconds exposures for the NIR observations. The first 8 spectra, however, were reported to be ``useless due to misconfiguration of the instrument'' by the CAHA staff, although that did not seem to have a discernible impact on the results, so those 8 initial spectra were kept in the sample.

In addition to these new observations, the CAHA database was queried for all of the publicly available CARMENES data since its first light in 2016. From the database query, a first filter was applied to only keep objects with spectral type O, B or A. Then, the sample was limited to objects classified as ``stars'' or ``'high proper motion stars'' to avoid having to deal with object types that may potentially exhibit complex spectral features potentially altering our normalization procedure (Section \ref{sec_normalisation}). This was done by cross checking the target names from the observation header with the SIMBAD catalogue \citep{simbad}. 

From the sample of stars, a further cut was applied to remove those observations that had adverse weather conditions such as high humidity (> 80\%) or high airmass (> 2). Out of the remaining observations, a simple test was devised to remove any spectra that did not have a strong enough signal. To do that, a continuum window was selected next to one of the \ce{CO2} lines. The range defining the line is $1.57817\ \mu \text{m} < \lambda < 1.57827\ \mu \text{m}$, while the range defining the continuum is $1.57828\ \mu \text{m} < \lambda < 1.57860\ \mu \text{m}$. The signal strength was quantified by calculating the difference between the mean of the continuum window and the depth of the line and dividing it by the continuum standard deviation. If this value was below 20, the spectra was deemed ``too noisy'' and removed from the sample. The limiting value was determined upon inspection of all of the remaining observations which verified that the spectra whose signal strength fell below this value were indeed noisy and devoid of any significant signal. With this cut, the final CARMENES sample is determined, containing 600 observations spanning the range July 2016 to December 2023, which includes the aforementioned observations carried out on the weather balloon night. 

\subsubsection{CARMENES technical specifications}\label{carmenesdata}
CARMENES has two separate échelle spectrographs, one covering the visible wavelength range, from $0.55 - 1.05\ \mu$m, and the other covering the near infrared range, from $0.95 - 1.7\ \mu$m. Only the latter is of interest to this work, as it is in the near infrared that prominent \ce{CO2} and \ce{CH4} telluric lines are found. The resolving power for CARMENES in the NIR is reported to be $R = 80400$ \citep{quirrenbach18} and is obtained from the average of measurements of unresolved lines of a hollow cathode lamp \citep{quirrenbach16}. 

CARMENES data products come in pairs, one corresponding to the data obtained by the science fiber A and the other corresponding to the data obtained by the calibration fiber B, which in most cases is pointed to the sky. Here, only data coming from fiber A is used. The science data is divided in orders, each covering a portion of the whole wavelength range. There are 28 orders in total, and for this analysis only 6 of them were used, namely the ones that contained prominent absorption lines associated with the molecules relevant to this study. Table \ref{tab_carmenes_orders} gives a summary of which orders were used, their wavelength range and which molecules contain prominent lines in this range.

\begin{table}
    \centering
    \begin{tabular}{ccc}
        \hline 
        Order & Wavelength range ($\mu$m) & Molecular lines \\
        \hline
         12 & $1.16 - 1.19$ & \ce{H2O} \\
         16 & $1.26 - 1.28$ & \ce{O2} \\
         25 & $1.55 - 1.58$ & \ce{CO2} \\
         26 & $1.59 - 1.62$ & \ce{CO2} \\
         27 & $1.64 - 1.66$ & \ce{CH4} \\
         28 & $1.68 - 1.71$ & \ce{CH4}, \ce{H2O} \\
    \end{tabular}
    \caption{Summary of CARMENES orders used in the analysis, along with the wavelength range they cover and the prominent molecular lines found in this range for the molecules relevant to this study.}
    \label{tab_carmenes_orders}
\end{table}

\subsubsection{Normalisation process}\label{sec_normalisation}
The model spectra described in Section \ref{sec_methodology} is computed normalised to 1, but that is not the case for the observational spectra, as the measured flux level is affected by a number of sources, such as instrumental systematics and physical effects like stellar variability. Therefore, both the model spectra and the observational spectra need to be normalised before they can be compared to each other.

The normalisation process involves creating a ``normalisation mask'' that is used to determine what is a spectral line and what is part of the continuum, and then running a median filter only on the continuum points. The normalisation mask is created from a dummy model spectra containing arbitrary values for the molecular abundances of \ce{CO2} and \ce{H2O}, listed in Table \ref{tab_algorithm_rules}. These arbitrary values are chosen so that the lines are not saturated and the line density is not so large such that the continuum cannot be easily identified. A first median filter is run through the whole spectra (including the lines) to bring the continuum level down to zero. Because the dummy model spectra is expected to contain no noise, any variation below a certain level indicates a line's presence. To quantify that level, we use the median absolute deviation (MAD) of the spectra, and everything below $-1 \times$ MAD is considered a line.

This mask is then used to normalise the observational spectra and the model spectra in all steps of the MCMC. The normalisation is done by running a second median filter only on the masked dataset, which should include only the continuum points, and then linearly interpolating it to fill in the gaps left by the lines. Finally, the normalised spectra is obtained by simply dividing the original spectra by the interpolated median filter. 

For each median filter, a window size must be chosen. For CARMENES, the spectra is divided in orders, so there is a different window size value for each order. These are also shown in Table \ref{tab_algorithm_rules}.

All of these parameters (molecular abundances for dummy model spectra, rule to define what is a line, median filter window sizes), however, did not prove to significantly alter the results when within reasonable limits.

\subsubsection{Emission lines}\label{sec_methodology_em_lines}
The model spectra computed here do not include emission lines, so these must be removed from the observational spectra to avoid any issues. This is done by using telluric emission spectra from the ESO Sky Model Calculator \citep{noll2012,jones2013}. Details of this data set are given in Section \ref{sec_esodata}. The emission spectra is linearly interpolated to the same wavelength sampling as the observational spectra, and then a mask is created to identify the positions of the emission lines. This is done by defining a flux level above which everything is considered to be an emission line. The points considered to be emission lines are then masked out of the spectra. By being too conservative in defining the flux level limit, some emission lines that could cause trouble in the analysis may not be removed, but by being too lenient, too much of the spectra might end up being removed, along with some absorption lines in observations where emissions lines may not even be present. There is one minimum flux level defined for each CARMENES order used, which were determined by visual inspection of the emission lines, and they are listed in Table \ref{tab_algorithm_rules}.

\subsubsection{Deep and saturated absorption lines}
A ``deep line cut'' was applied to remove points below a certain transmission threshold from the calculation of the log likelihood function. The threshold chosen was 0.2. The reason for this cut is because the signal from deep lines may be comparable to the noise in the data, plus saturated lines are problematic because information is lost. This rule is also included in Table \ref{tab_algorithm_rules}.

\subsection{Atmospheric data}\label{sec_atmdata}
To properly model the spectral lines, we require temperature, pressure and molecular abundance values as a function of height, which collectively are referred to as the ``atmospheric profile''. The molecules selected to be included in the model were \ce{CO2}, \ce{CH4}, \ce{H2O}, \ce{O2} and \ce{N2}. The first two are the GHGs of interest for this study and the other three are molecules abundantly present in the Earth's atmosphere.

\subsubsection{Literature atmospheric profiles}\label{sec_atmprofs}
The Michelson Interferometer for Passive Atmospheric Sounding \citep[MIPAS;][]{fischer08}\footnote{\href{http://eodg.atm.ox.ac.uk/RFM/atm/}{eodg.atm.ox.ac.uk/RFM/atm/}} was a mid-infrared emission spectrometer onboard the ENVISAT \citep{louet1999} satellite. The atmospheric profiles provided by MIPAS have all of the aforementioned parameters, but they are only computed for general locations such as mid-latitude (day and night), polar winter/summer and equatorial day-time rather than for specific locations, plus there is no time information on the profiles. Here, the equatorial profile was used.

On the other hand, the Global Data Assimilation System \citep[GDAS;]{kanamitsu1989,derber1991}\footnote{\href{https://ftp.eso.org/pub/dfs/pipelines/skytools/molecfit/gdas/}{ftp.eso.org/pub/dfs/pipelines/skytools/molecfit/gdas/}}, provided by the US National Oceanic and Atmospheric Administration (NOAA), has profiles for specific locations around the globe, the aforementioned Calar Alto Observatory being one of them. Additionally, their profiles are computed every 3h, so we can get conditions very close to the time of observation. However, the GDAS profiles only contain information on the molecular abundance of water (i.e., the humidity), so the MIPAS profile is still required to obtain information on the other molecules. Whenever employing GDAS profiles, the two profiles closest in time to the observation are linearly interpolated to obtain the values at the time of observation.

Both the MIPAS and the GDAS profiles come from actual measurements of the atmosphere. MIPAS used to observe different atmospheric levels at a line of sight that penetrated down to a minimum altitude, and the profiles were obtained from a forward retrieval model \citep{fischer08}, whereas the GDAS profiles combine different types of observations such as surface observations, balloon data, aircraft reports and satellite observations \citep{rodell2004}.

The GDAS profiles span an irregular grid from $\sim 0.1-26$ km, whereas the most recent MIPAS profiles go from $0-200$ km in constant steps of 1 km. However, after around 80 km, the pressure is so low that they are below the limits of the computed cross-sections, so the MIPAS profile is clipped and kept only up to there. 

\subsubsection{Weather balloon launch}\label{sec_weather_balloon}
To complement these data sets, a new atmospheric profile was measured specifically for this work, using a weather balloon. The balloon launch was carried out on the same night as the spectroscopic observations described in Section \ref{carmenesdata}. All of the materials necessary for the launch were purchased from Stratoflights\footnote{\href{https://www.stratoflights.com/en/}{www.stratoflights.com/en/}}, who provide a complete kit that contains all of the essentials to carry out such an experiment. The electronics responsible for carrying out the measurements were the DataloggerSTRATO4, which could measure the outside temperature, pressure and humidity. It also kept track of time and location (latitude, longitude and altitude). Measurements were taken with a 2 second cadence, which results in a quite fine height grid, so to keep computational time manageable when using the balloon profile, the original height grid of the datalogger was linearly interpolated to a new grid of 50 height points. This new height grid is explained in more detail in the next section.

\begin{figure}
    \centering
    \includegraphics[width=\columnwidth]{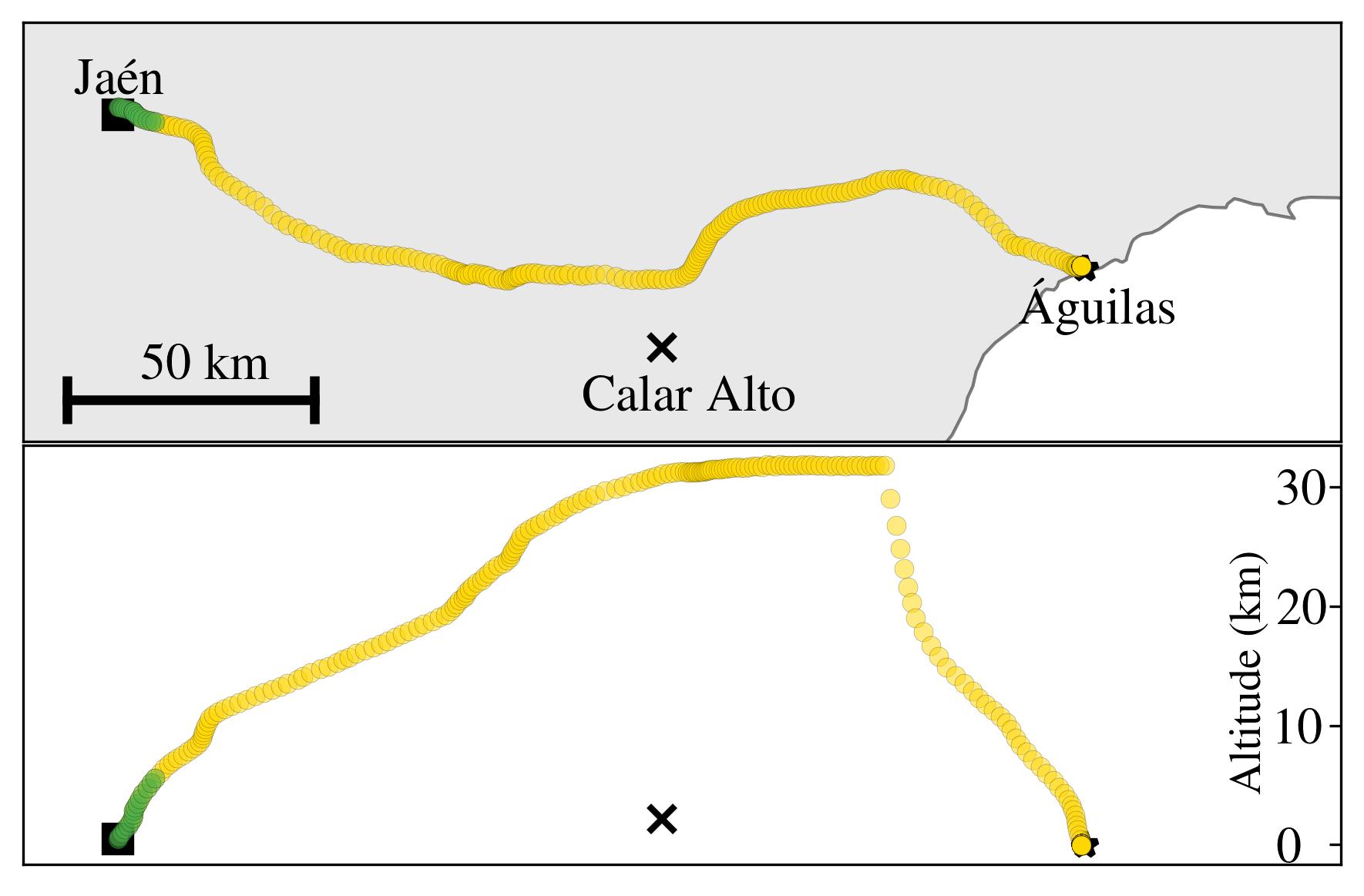}
    \caption{Top: flight path of the weather balloon. Locations of the Calar Alto Observatory, the city of Jaén (launch site) and the municipality of Águilas (crash site) are indicated by the black cross, the black square and the black star, respectively. Map scale is shown on the bottom left corner. Bottom: altitude variation throughout the flight. Symbols indicate the altitudes of the aforementioned sites. The green circles correspond to the points prior to the datalogger malfunction, while the yellow circles are the points after that. For visualisation purposes, the data set was plotted with a 100s cadence instead of 2s.}
    \label{fig_balloon_flight_path}
\end{figure}

Taking all of the logistics and safety measurements into account, which included flight path simulations\footnote{\href{https://www.stratoflights.com/en/tutorial/weather-balloon-tools/predicting-the-flight-path/}{www.stratoflights.com/en/tutorial/weather-balloon-tools/predicting-the-flight-path/}}, the selected launch site was the city of Jaén, around 125 km away from the Calar Alto Observatory. Even with these extra precautions, and despite being more than 100 km from the nearest shore, the balloon almost landed in the ocean, in the municipality of Águilas, as can be seen from Figure \ref{fig_balloon_flight_path}, which shows the flight path of the weather balloon.

Due to an unknown electronics malfunction that happened during the flight, the external sensor stopped recording data around 30 minutes after launch, when it had risen to a height of around 6 km. The trajectory of the weather balloon prior to the electronics malfunction is represented by the green circles in Figure \ref{fig_balloon_flight_path}.

\subsubsection{Combining the atmospheric profiles}\label{sec_combprofs}
In order to get the most accurate atmospheric profile possible, all the available profiles are employed. That includes site data, balloon data, GDAS data and MIPAS data. The site data is collected by the observatory's weather station and is included in the file header of each observed spectrum. The site data, therefore, are actual measurements of pressure, temperature and humidity (but no other molecular abundances) taken at the time of the observation at the height of the telescope. The combined atmospheric profile consists of site data plus whatever profiles we choose to attach on top of it. To ensure a smooth transition between the site data and the subsequent profile, an approach similar to that described in \cite{noll_eso_skymod_doc} was applied, which adjusts the contribution of the site data by a scaling factor for heights $h$ below a critical height $h_\text{crit}$. The corrected profile value is calculated as (in the equations below, $\nu$ denotes the site or profile measurements, as indicated by the subscript):

\begin{equation}\label{ballooncorr}
    \nu _\text{profile,corr} (h) = c(h) \times \nu_\text{profile} \quad ,
\end{equation}
where the correction factor $c(h)$ is given by:
\begin{equation}\label{corrfact}
    c(h) =
\begin{cases}
c(h_\text{site}), & h < h_\text{site} \\
(1-b)\frac{h}{h_\text{crit}} + b & h_\text{site} < h < h_\text{crit} \\
1, & h > h_\text{crit}
\end{cases}
\quad ,
\end{equation}
where
\begin{equation}
    b = \frac{c(h_\text{site})h_\text{crit} - h_\text{site}}{h_\text{crit} - h_\text{site}}
\end{equation}
and
\begin{equation}
    c(h_\text{site}) = \frac{\nu _\text{site}}{\nu _\text{profile} (h_\text{site})}
\end{equation}

For Cerro Paranal, an inspection of the GDAS data showed that there is a gradual reversal of wind direction from sea level up to around 5 km, beyond which it remains constant, as reported in the documentation for the Cerro Paranal Advanced Sky Model \citep{noll_eso_skymod_doc}. Due to this feature in the atmospheric data, it is argued in the documentation that it is safe to assume that above this critical altitude the air properties are no longer significantly correlated with the local site measurements and are described by the GDAS data only, so here we employ $h_\text{crit} = 5$ km as well. Using the corrected profile values to merge site and profile data is referred to as the ``complex'' site merging approach. In contrast, simply plugging the profile values on top of the site values is referred to as the ``simple'' site merging approach.

Since the MIPAS profile is computed for a generic latitude and has no time information, it is deemed as ``less representative'' of the conditions at the time of observation. Between the GDAS and the balloon profiles, either could be argued as the best option, so both options were explored. This process is described in Section \ref{sec_atmprofile_study}. 

To combine different atmospheric profiles, a new height grid was constructed. This new grid goes from the site height up to the highest point available in all profiles, which in this case is 80 km. The points in the new grid are evenly spaced in $\log{(h)}$, resulting in a smaller spacing for the points closer to the ground, which have a larger influence on the model because pressure and abundances are larger there. The bottom profiles (i.e., the balloon and GDAS profiles) are only interpolated up to their maximum height values to avoid having to extrapolate. In the transition points between profiles, a weighted mean is applied that quantifies the contribution of each profile in order to smoothly go from one to the other. For the Cerro Paranal Advanced Sky Model, they use 80\%, 60\%, 40\% and 20\% as the contributions of the bottom profile in the transition points between the bottom and top profiles (the contribution decreases as the altitude increases). Here, the same weights are used.

\begin{figure}
    \centering
    \includegraphics[width=\columnwidth]{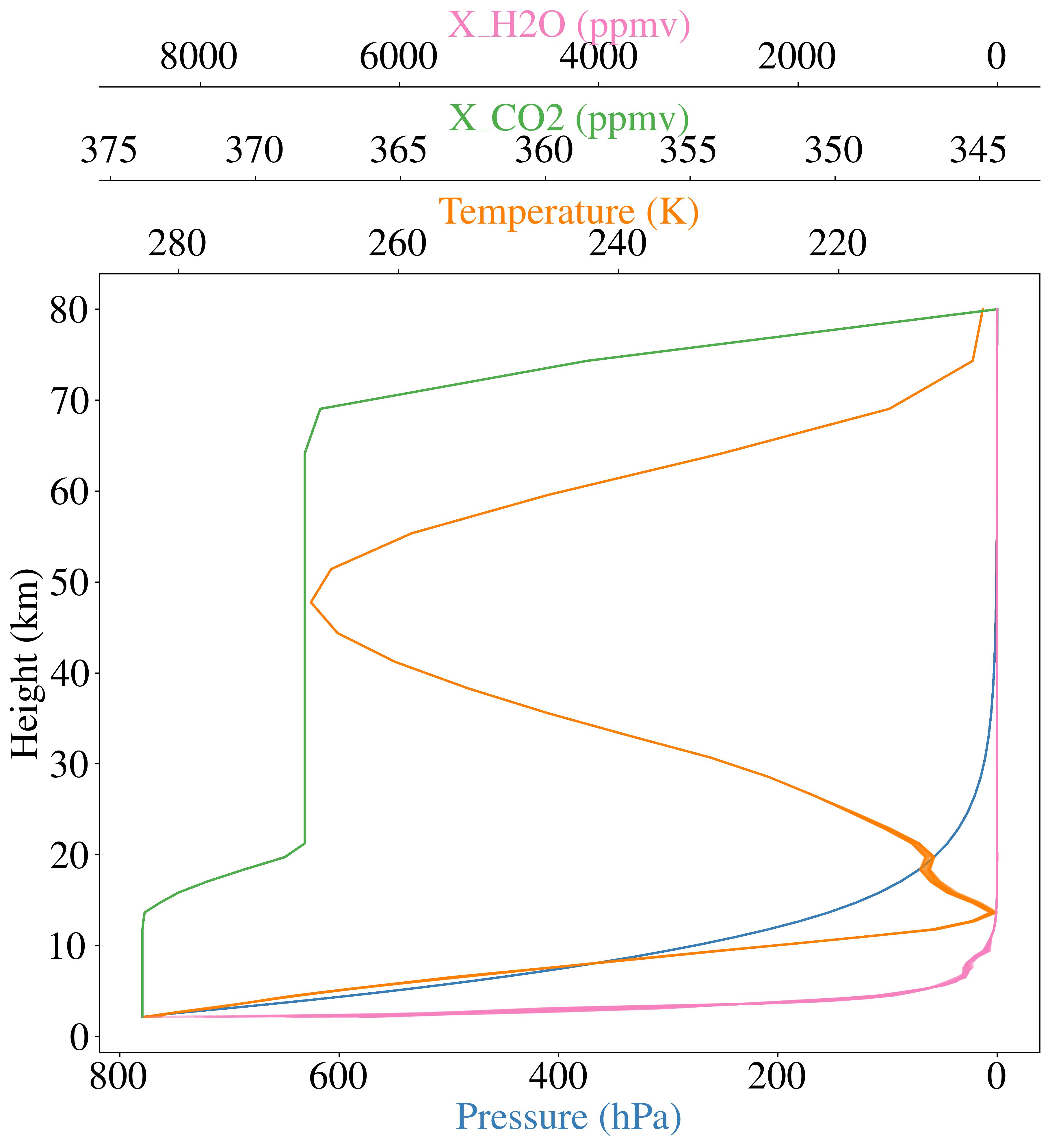}
    \caption{Atmospheric profiles for each of the observations taken with CARMENES on the weather balloon night. Pressure, temperature, \ce{CO2} abundance and \ce{H2O} abundance are shown as a function of height in blue, orange, green and pink, respectively. The profiles are obtained by combining site data, GDAS data and MIPAS data using the simple site merging approach and smoothing out the transition between profiles.}
    \label{fig_atmospheric_profiles}
\end{figure}

The humidity profile was also constructed as explained above. However, because only the MIPAS profile contains information about the other molecules, their profiles were obtained by simply linearly interpolating the MIPAS profile to the new height grid. Figure \ref{fig_atmospheric_profiles} shows the resulting atmospheric profiles for all of the CARMENES observations on the night of the balloon launch for the case where only the GDAS and the MIPAS profiles were used and combined using the simple site merging and smooth profile transition (i.e., the two profiles were merged using a weighted mean as opposed to just plugging them on top of each other).

For the analysis described in Section \ref{sec_atmprofile_study}, different combinations of the GDAS and MIPAS profiles were employed alongside the weather balloon profile measured for this work. There were four scenarios: simple merging of site data and profile data, no balloon data; complex merging of site data and profile data, no balloon data; simple merging of site data and profile data, with balloon data; and complex merging of site data and profile data, with balloon data.

\subsubsection{GGG2020 atmospheric profiles}\label{sec_ggg2020profs}
The GGG pipeline is used by the TCCON to process their spectra and compute the total column abundances of the relevant gases \citep{wunch2011}. The current version of the pipeline is GGG2020 \citep{laughner23a} and one of its applications is an algorithm that allows the user to compute a priori profiles for several gases. This algorithm is called ``\texttt{ginput}'' and is publicly available on GitHub \citep{laughner2021}.

For the main gases, such as \ce{CO2} and \ce{CH4}, the basic process to compute profiles is similar. Their mole fractions are tied to the monthly average measurements in whole-air flasks from the Mauna Loa (MLO) and American Samoa (SMO) sites from the GML network, and the underlying assumption is that the concentration of the gases at a certain time and place is a function of the concentration at the aforementioned sites and the time of transport between them and the profile location, including chemistry occurring during stratospheric transport \citep{laughner23a}.

\texttt{ginput} also relies on meteorological parameters, and here the Goddard Earth Observing System Forward Product (GEOS FP)\footnote{\href{https://gmao.gsfc.nasa.gov/GMAO_products/}{https://gmao.gsfc.nasa.gov/GMAO\textunderscore products/}} reanalysis product was used.

The code was tailored to include a custom location for the Calar Alto Observatory. The resulting atmospheric profiles covered a height range of around 1-77 km.

The MLO and SMO data used to create the GGG2020 profiles only goes until 2018 and will not be updated until the next GGG release to ensure consistent priors are created in the current version \citep{laughner23a}. Therefore, when creating profiles after 2018, it is necessary to extrapolate the MLO and SMO records. To do that, \texttt{ginput} fits a function $f(t)$ to the last $n$ years of the MLO and SMO records, then calculates the average seasonal cycle over these years as the anomaly relative to $f(t)$ and extrapolates to the desired date using $f(t)$ as the baseline and applying the average seasonal cycle on top of it \citep{laughner23a}. For \ce{CO2}, the number of years is $n = 10$, whereas for \ce{CH4} $n = 5$, and the functions used are, respectively: 

\begin{equation}\label{eq_CO2_fit_func}
    f(t) = c_0 e^{c_1 t}
\end{equation}

\begin{equation}\label{eq_CH4_fit_func}
    f(t) = c_0 + c_1 t + c_2 t^2
\end{equation}

It should be noted that extrapolating the MLO and SMO dry mole fractions (DMFs) can introduce errors in the retrievals, for example if the profile shape is wrong when dealing with an El Niño year. \cite{laughner23a} estimate that the current error in the current MLO and SMO DMFs due to extrapolation is around 0.25\% for \ce{CO2} and 0.6\% for \ce{CH4}. At present, these errors are not propagated into the uncertainties reported in this work.

\begin{figure}
    \centering
    \includegraphics[width=\columnwidth]{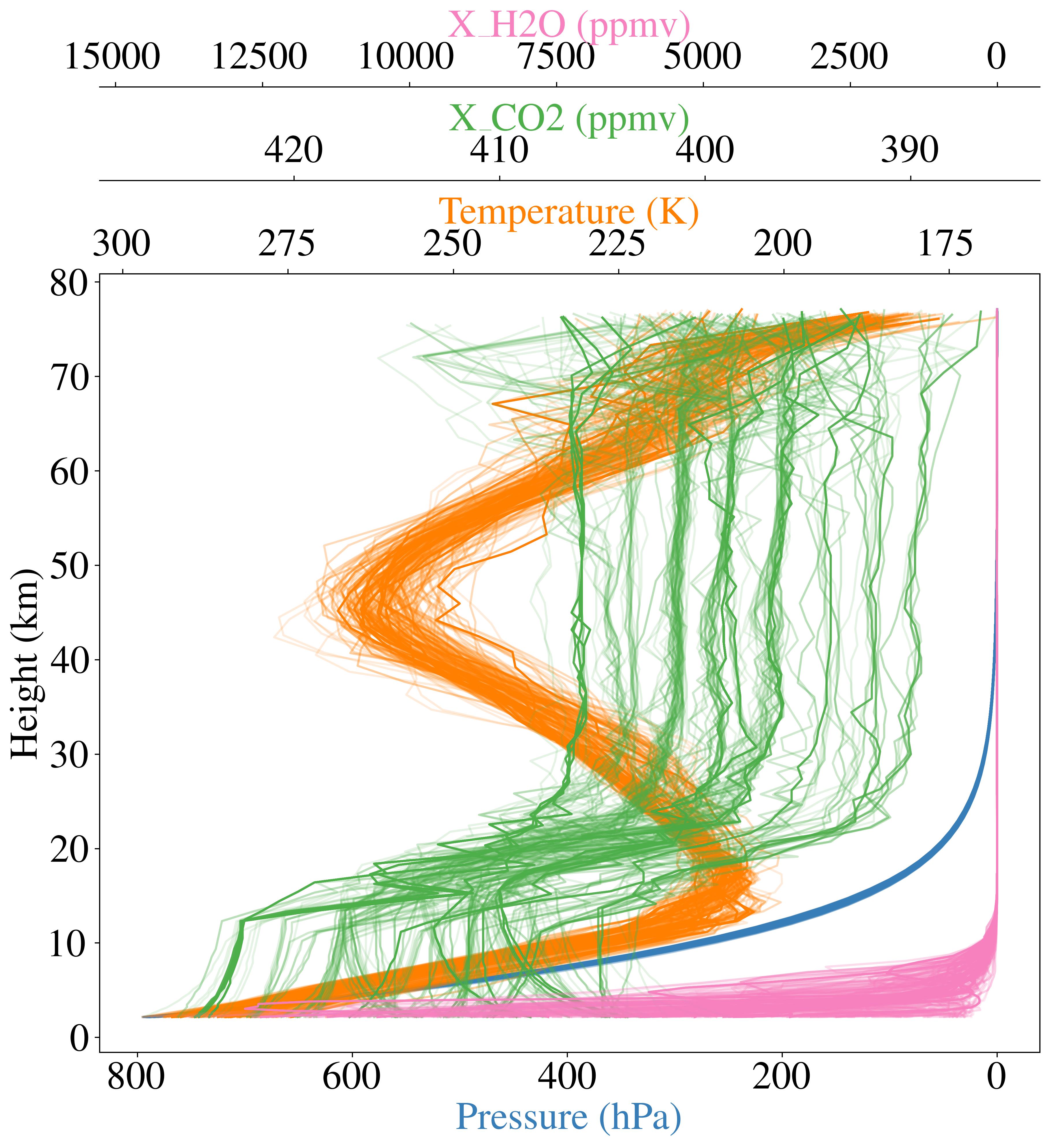}
    \caption{Atmospheric profiles for the entire CARMENES data set, generated with \texttt{ginput}. Pressure, temperature, \ce{CO2} abundance and \ce{H2O} abundance are shown as a function of height in blue, orange, green and pink, respectively.} 
    \label{fig_atmospheric_profiles_ggg2020}
\end{figure}

The collection of atmospheric profiles generated with \texttt{ginput} for each of the CARMENES observations that make up our dataset is shown in Figure \ref{fig_atmospheric_profiles_ggg2020}.
Unlike with the GDAS and MIPAS profiles, knowledge of the site measurements from the observatory’s weather station is not incorporated when using the GGG2020 profiles.

\subsection{CAMS reanalysis model}
Since the measurements of the abundance of greenhouse gases shown here are taken during the night, they cannot be directly compared to the measurements from networks that rely on sunlight. In addition, ground-based networks such as the TCCON do not have any sites close to the Calar Alto Observatory, and care must be taken when comparing ground-based measurements with measurements from collocating satellite overpasses if we were to use OCO-2, OCO-3 or GOSAT-2 as reference. 

The solution is to use a reanalysis model as a benchmark for our results. Here, we used the CAMS EGG4 reanalysis model \citep{inness2019, agusti-panareda2023}. This dataset is produced by the European Centre for Medium-Range Weather Forecasts (ECMWF) and can be accessed from the CAMS Atmosphere Data Store \citep[ADS;][]{CAMSEGG4}, with data available from 2003 to 2020.

Among the parameters included in this reanalysis model, the ones relevant for this work are the column-averaged molar fraction of \ce{CO2} and \ce{CH4}, as well as the ``ground-level abundance'' for these two molecules plus \ce{H2O}, which here means the abundance at the height of the Calar Alto Observatory. The surface pressure is a parameter given in the header of observations taken at that observatory, and in the time period covered by the CARMENES observations employed here, the surface pressure in Calar Alto ranged from $\sim 770-790$~hPa. The EGG4 data can be retrieved based on pressure levels, and the closest one to the surface pressures measured in Calar Alto is 800~hPa, so that is what was used. 

The multi-level molecular abundances are given in units of kilogram of gas per kilogram of dry air. To convert those to parts per million (ppm), they must be scaled by the ratio of the molar mass of dry air and the molar mass of the gas and then multiplied by $10^6$ to get the value in ppm. The molar mass of dry air was taken to be $M_\text{air} = 28.96546$~g~mol$^{-1}$ \citep{picard2007}, while the molar masses of \ce{CO2}, \ce{CH4} and \ce{H2O} are $M_\text{\ce{CO2}} = 44.0095$~g~mol$^{-1}$, $M_\text{\ce{CH4}} = 16.0425$~g~mol$^{-1}$ and $M_\text{\ce{H2O}} = 18.0153$~g~mol$^{-1}$, respectively, all taken from the NIST database \citep{nist}\footnote{\href{https://webbook.nist.gov/chemistry/}{https://webbook.nist.gov/chemistry/}}.

From the resolution of the CAMS reanalysis, the closest point to the Calar Alto Observatory had the coordinates $37.20^\circ$N and $2.55^\circ$W, which is within 3 km of the observatory. The CAMS reanalysis model has a time resolution of 3h. It should be noted that no uncertainties are reported for the EGG4 parameters.

\subsection{ESO Sky Model data}\label{sec_esodata}
As mentioned in Section \ref{sec_methodology_em_lines}, removing emission lines from the observational spectra is a necessary step in the modelling process, as they are not included in \textit{Astroclimes}. To do that, the ESO Sky Model Calculator (SKYCALC)\footnote{\href{https://www.eso.org/sci/software/pipelines/skytools/skymodel}{https://www.eso.org/sci/software/pipelines/skytools/skymodel}} was used. SKYCALC is a web application based on the Cerro Paranal ASM that allows one to generate both telluric transmission and emission spectra. 

Among the parameters that can be changed in SKYCALC, there are observatory height, airmass, season, period of night, precipitable water vapour (PWV), wavelength range and binning, and whether to convolve it with a line spread function to broaden the spectra or not. The default values are set to an observatory height of 2640 m (Cerro Paranal), airmass of 1, season and period of night are entire year and entire night, PWV~=~2.5~mm, a logarithmic binning of $\frac{\lambda}{\Delta \lambda} = 20000$ and no broadening. The default wavelength range is $1-2~\mu$m, but we limited it to $1-1.8~\mu$m since there was no need to go all the way to $2~\mu$m. 

The sky emission model used had the default setup and the same template was used for all observations because the emission model is simply used to locate the position of the lines. The details on how the emission spectra were used in the modelling process are described in Section \ref{sec_methodology}.

Transmission spectra from SKYCALC were also employed in this work. Transmission telluric spectra were generated for each of the 13 different available PWV values in SKYCALC: PWV = 0.05, 0.10, 0.25, 0.50, 1.00, 1.50, 2.50, 3.50, 5.00, 7.50, 10.00, 20.00 and 30.00, all in mm. Every other parameter was kept as default, except the resolution, which was chosen to be $R=300000$. These spectra were used as reference to test the capabilities of \textit{Astroclimes}, a process which is described in Section \ref{sec_esocomp}.

\section{Results and Discussion}\label{sec_results}
The resulting synthetic transmission model from \textit{Astroclimes} contains telluric lines associated with the aforementioned selected molecules, namely \ce{CO2}, \ce{CH4}, \ce{H2O}, \ce{O2} and \ce{N2}. The fitting algorithm developed aimed to find the abundance values from these molecules that resulted in the best fit when compared to observational data. Because \ce{N2} does not contain strong spectral features in the wavelength region used here \citep{smette15}, its abundance was kept fixed, so the free parameters included are the abundances of \ce{CO2}, \ce{CH4}, \ce{H2O}, \ce{O2}.

The molecular abundances are included in the modelling process when the transmission at each atmospheric layer is calculated. Changing the molecular abundances in each layer individually would require too much computational time, plus our measurement approach is not suitable to constrain the altitude dependence of molecular abundances. Instead, free variables are parametrised by their ``ground level abundance'', which is the abundance given by their respective profiles at the height of the observatory. This ground level abundance is allowed to vary, and the rest of the profile is scaled accordingly, i.e. the molecular abundance profiles are being multiplied by a scaling factor.

The abundance profiles taken from the GGG2020 profiles are given in DMF, but for practical reasons, the abundances are handled in number density units throughout the fitting process. To convert the abundance of a gas from DMFs $c_\text{gas}$ to number densities $n_\text{gas}$, a rearranged version of equation (9) from \cite{laughner23a} was used:

\begin{equation}\label{eq_dmf}
    c_\text{gas} = \frac{n_\text{gas}}{n_\text{ideal}} (1 + c_\text{\ce{H2O}})
\end{equation}

In the equation above, $n_\text{ideal}$ is the ideal gas number density, given by equation (\ref{eq_idealgaslaw}), and $c_\text{\ce{H2O}}$ is the DMF of water, obtained from the prior \ce{H2O} profile. To convert it back to DMFs, equation (\ref{eq_dmf}) is first rearranged to get the posterior DMF distribution of \ce{H2O}, and then the DMF of the other molecules can be calculated, again using equation (\ref{eq_dmf}). 




The column abundance of a gas can be parametrised by its column-averaged DMF, denoted $X_\text{gas}$ and defined as the ratio of the column abundances of said gas to the column abundance of dry air:

\begin{equation}\label{eq_xco2}
    X_\text{gas} = \frac{\int _{0}^{\infty} n_\text{gas} (z) \text{dz}}{\int _{0}^{\infty} n_\text{air} (z) \text{dz}} \quad ,
\end{equation}
where $n_\text{gas} (z)$ is the altitude dependent number density of the gas, that is, the number of gas molecules per cubic meter, and $n_\text{air} (z)$ is the same quantity for dry air. Equation (\ref{eq_xco2}) can be rewritten in terms of the number density of oxygen since the dry air mole fraction of \ce{O2} in the atmosphere, $\text{DMF}_\text{\ce{O2}}$, is a known quantity and essentially constant \citep{crisp2015}, so we get:

\begin{equation}\label{eq_xco2_o2}
    X_\text{\ce{CO2}} = \text{DMF}_\text{\ce{O2}} \times \frac{\int _{0}^{\infty} n_\text{\ce{CO2}} (z) \text{dz}}{\int _{0}^{\infty} n_\text{\ce{O2}} (z) \text{dz}} \quad ,
\end{equation}

The DMF of \ce{O2} has been reported to be 209500 ppm in papers related to GOSAT \citep{morino2011} and TCCON \citep{laughner23a}, but 209350 ppm in documentation related to OCO-2 and OCO-3 \citep{crisp2015,crisp2020_ocoretalg}. Here, $\text{DMF}_\text{\ce{O2}} = 209500$ ppm is used, since the generated atmospheric profiles come from the TCCON GGG2020 pipeline.

For water, another useful quantity to express abundance is the ``precipitable water vapour'' (PWV), which is a measure of the total amount of water vapour contained in an air column above a certain site and it is usually expressed as the height of liquid water in mm that this water vapour would correspond to. This can be calculated by the following expression \citep{smette15}:

\begin{equation}\label{eq_pwv}
    PWV = \frac{M_{\ce{H2O}}}{\rho _{\ce{H2O}} R} \int _{z_0}^{\infty} \frac{x_{\ce{H2O}}(z) P(z)}{T(z)} \text{dz} \quad ,
\end{equation}
where $M_{\ce{H2O}}$ is the mole mass of water, $\rho _{\ce{H2O}} \approx 10^3$ kg/m$^3$ is the density of liquid water, $R = 8.31446\ \frac{\text{J}}{\text{mol} \text{K}}$ is the gas constant and $z_0$ is the height at which the PWV is being calculated. Numerically, this is done by employing the \texttt{numpy.trapz} function.

The free parameters' posterior distributions are inferred by running an MCMC, which is done with the help of \texttt{emcee} \citep{emcee}. To quantify the resemblance between the model and the observational data, the log likelihood function given by equation (\ref{eq_loglike}) was employed, which comes from \cite{brogiline}, as explained in Section \ref{sec_comp_mod_data}. 

In every application of an MCMC shown here, the same convergence criteria was used. One of the available statistical parameters in \texttt{emcee} is the autocorrelation time, which basically determines the number of steps needed for the chain to ``forget'' where it started (because the samples in the chain are not independent) \citep{emcee}. An MCMC run would be deemed ``converged'' if the autocorrelation time of each walker was less than the current step number divided by 100 and if the difference between the previous autocorrelation time and the current one was less than 1\%. Tests for convergence were carried out every 100 steps. If the convergence criteria was not satisfied, then the MCMC would simply run until the maximum specified number of steps. 

For all MCMC runs carried out here, the same setup was used, which had 10 walkers and 5000 maximum steps. The posterior distribution is obtained using a burn-in of half the total amount of steps, and the results are calculated as the median of the posterior distribution, with the uncertainties being its standard deviation.

\subsection{Fit to the ESO Sky Model}\label{sec_esocomp}
Both \textit{Molecfit} and the Cerro Paranal ASM are well-established tools in the field of astronomy \citep{smette15,kausch2015,noll2012,jones2013}. These were developed in parallel by ESO\footnote{\href{https://www.eso.org/sci/software/pipelines/skytools/}{https://www.eso.org/sci/software/pipelines/skytools/}} and use a very similar approach to generate their transmission spectra, which includes a similar way to obtain and combine atmospheric profiles, using the HITRAN line list database, and relying on the widely known line-by-line- radiative transfer model (LBLRTM) by \cite{clough2005}. Therefore, the SKYCALC models (web application of the Cerro Paranal ASM) were used as a benchmark test for \textit{Astroclimes}.

An MCMC was run, with the SKYCALC transmission spectra described in Section \ref{sec_esodata} used as the ``data'' for which the best-fit model had to be found. Starting with the values from the MIPAS profile, the goal was to see if \textit{Astroclimes} would be able to reproduce the reported PWV values from SKYCALC. As such, the setup was as close to the one used by the ASM as possible and is described in the paragraph below. When changing the PWV values of the ESO models, the abundances of the other molecules should remain unchanged (Noll, private communication).

The atmospheric profile used for this analysis is a combination of the GDAS and MIPAS profiles, plus site data. Site data was collected from January 5th 2023, arbitrarily chosen. The MIPAS profile used was the equatorial one, as this is the recommended one for Cerro Paranal according to the ASM documentation \citep{noll_eso_skymod_doc}. The critical height chosen was 5 km, as this is the default for the ASM, and the merging of the GDAS and MIPAS profiles was done with four weights ranging from 80\%-20\%, again following the documentation \citep{noll_eso_skymod_doc}. Because of the arbitrary choice of site values and due to the fact that the ASM interpolates the MIPAS and GDAS profiles to an irregular grid for which the levels are not stated, the resulting atmospheric profile is slightly different than the one employed to compute the SKYCALC models.

\begin{table}
    \centering
    \begin{tabular}{cc}
        \hline
        ESO PWV (mm) & Fitted PWV (mm) \\
        \hline
        0.05 & $0.05030 \pm 0.00001$ \\
        0.10 & $0.09929 \pm 0.00002$ \\
        0.25 & $0.24626 \pm 0.00006$ \\
        0.50 & $0.4942 \pm 0.0001$ \\
        1.00 & $0.9981 \pm 0.0003$ \\
        1.50 & $1.5098 \pm 0.0006$ \\
        2.50 & $2.539 \pm 0.001$ \\
        3.50 & $3.564 \pm 0.002$ \\
        5.00 & $5.075 \pm 0.003$ \\
        7.50 & $7.574 \pm 0.004$ \\
        10.00 & $10.048 \pm 0.005$ \\
        20.00 & $20.86 \pm 0.01$ \\
        30.00 & $32.81 \pm 0.02$ \\
    \end{tabular}
    \caption{Values and uncertainties obtained when running Astroclimes on the different SKYCALC models.}
    \label{tab_eso_results}
\end{table}

\textit{Astroclimes} managed to correctly reproduce the PWV values reported by SKYCALC in all cases except for PWV = 20mm, 30mm, as shown in Table \ref{tab_eso_results}, which are quite extreme cases that rarely occur in regions where telescopes are located. The reason for this discrepancy has not been thoroughly investigated, but it might be due to the fact that for such high humidity values most of the water lines are heavily saturated, so it may be hard to distinguish between different abundance values.

\subsection{Atmospheric profile study}\label{sec_atmprofile_study}
The choice of which atmospheric profile to use in the analysis is expected to play a crucial role on the results. Using the atmospheric profiles from the GGG2020 pipeline is the most straightforward option, as they provide all of the necessary parameters and don't require combining profiles from different sources. However, it is worthwhile to investigate how using different profiles influence the retrievals. 

\begin{figure}
    \centering
    \includegraphics[width=\columnwidth]{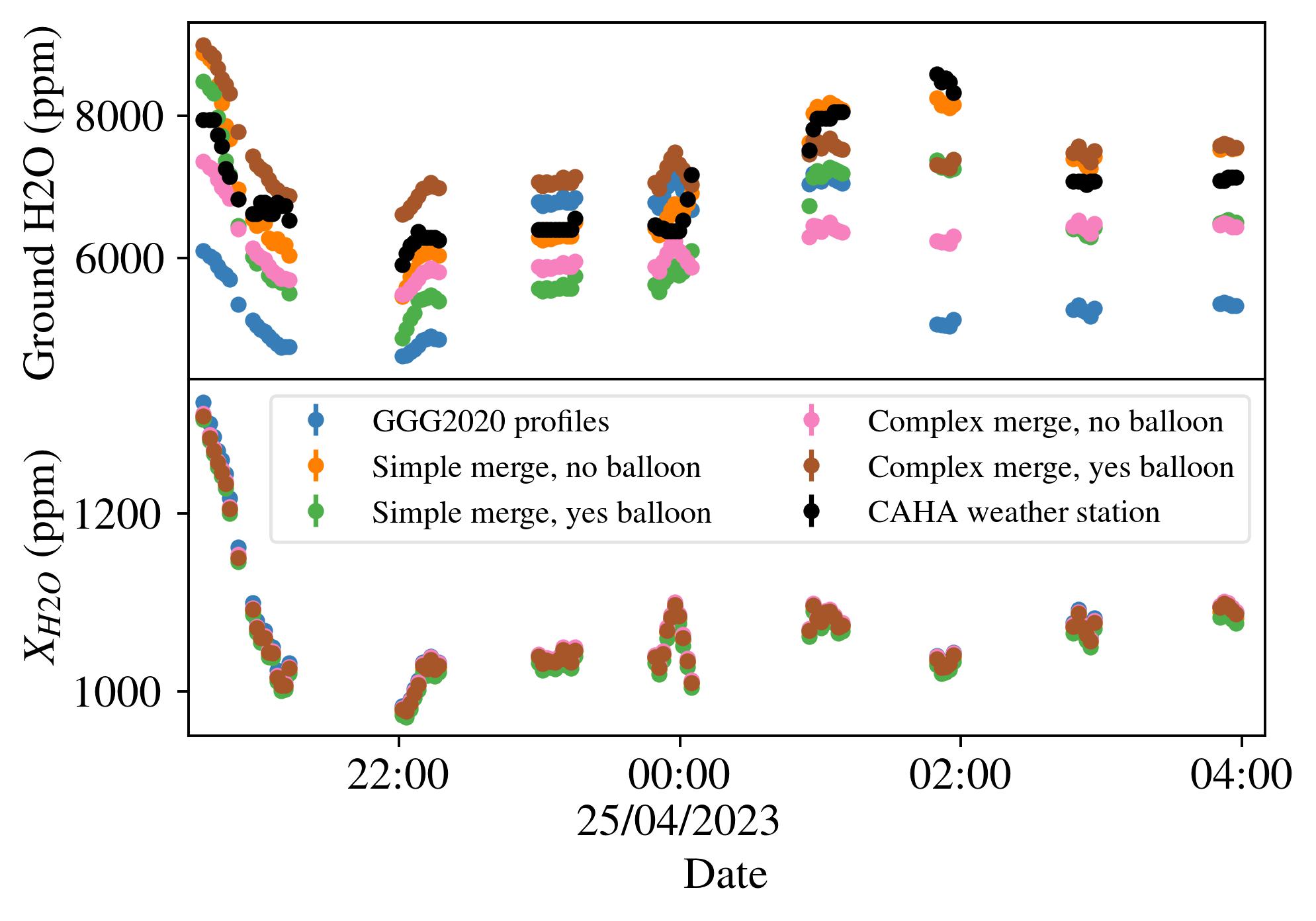}
    \caption{Top: ground level \ce{H2O} abundance at the night of April 24th 2023 as retrieved by the four different scenarios investigated in the atmospheric profile study carried out in this work: simple site merge, no balloon data (orange); simple site merge, with balloon data (green); complex site merge, no balloon data (pink); and complex site merge, with balloon data (brown). Plotted as well are the measurements from the observatory's weather station (black) and the results from the analysis using the GGG2020 atmospheric profiles instead (blue, more details on Section \ref{sec_carmenes_analysis}). Bottom: same as top panel, but for the column-averaged \ce{H2O} abundance as given by equation (\ref{eq_xco2_o2}), with no humidity measurements from the CAHA weather station.}
    \label{fig_h2o_atmtest}
\end{figure}

For this analysis, an MCMC was run on the CARMENES data from the weather balloon night using each of the four atmospheric profile combinations described in Section \ref{sec_atmprofs}. The resulting ground level and column-averaged ``dry air'' mole fractions of \ce{H2O} are shown in Figure \ref{fig_h2o_atmtest}.

From the top panel of Figure \ref{fig_h2o_atmtest}, we see that the choice of atmospheric profile has a non-negligible influence in our calculated ground level \ce{H2O} abundance, where ``ground level'' corresponds to the height at the observatory in question. However, its bottom panel shows that the total column-averaged DMF of \ce{H2O} is not very sensitive to the choice of atmospheric profile, evidenced by the retrievals of every scenario being clumped no top of each other. Therefore, without a reliable abundance profile, it is harder to achieve an accurate abundance measurement for a specific height.

\subsection{Retrievals from the CARMENES spectra and comparison to the CAMS reanalysis model}\label{sec_carmenes_analysis}
For the final analysis on the CARMENES sample, the atmospheric profiles used were the ones computed with the GGG2020 pipeline described in Section \ref{sec_ggg2020profs} and shown in Figure \ref{fig_atmospheric_profiles_ggg2020}. Despite the retrievals of the ground \ce{H2O} abundance using these profiles being the furthest match from the observatory's weather station measurements (Figure \ref{fig_h2o_atmtest}), the GGG2020 profiles are the only ones that contain all of the required parameters for the modelling process, such that there is no need to devise a strategy to combine atmospheric profiles from different sources.

Each of the observational spectra from CARMENES was analysed by running an MCMC as explained at the beginning of this section. After running this analysis for the first time on the CARMENES observations, it became clear that some problematic spectra still remained in the sample. This was evidenced by values of the retrieved ground \ce{O2} abundance that were very far from the mean dry mole fraction of \ce{O2} in the atmosphere, which has been reported to be 209500 ppm \citep{laughner23a} and, despite not being constant, does not vary by a lot \citep{bender1998, wunch2010}. Some of these cases were investigated in more detail and their spectra did exhibit features (either of stellar origin or due to instrumental or calibration errors) that could explain the discrepancy in the retrieved values, but a more robust analysis is needed to confirm that. Based on visual inspection of the distribution and the variability in the results, a further cut was implemented to the data and all observations that yielded \ce{O2} values outside of the range $209500 \pm 10000$ ppm were removed. Outliers among the other retrieved parameters were also examined for problematic features in their spectra and removed accordingly. With these cuts, the sample size went from 600 to 511 observations.

Some of the spectra also exhibited telluric lines that appeared to be shifted in wavelength, presumably due to a wavelength calibration error during the data processing. To identify such spectra, the model spectrum, which should have the correct line positions, was shifted in wavelength, and each time the cross-correlation between the model spectrum and the observational spectrum was calculated. The shift in wavelength went from $-25 \times 10^{-12}$ m to $+25 \times 10^{-12}$ m in steps of $1 \times 10^{-12}$ m. The expectation was that the highest cross-correlation would be achieved when there was no shift between observation and model. If that was not the case, then the observational spectrum was flagged as having a calibration error and removed from the sample, meaning that effectively the threshold to discard a spectrum was if it presented a shift larger than $1 \times 10^{-12}$ m. This brought our sample size down from 511 to 436 observations. These 436 observations comprise the final CARMENES sample and unless otherwise stated, all the results shown hereafter come from this sample.

\begin{figure}
    \centering
    \includegraphics[width=\columnwidth]{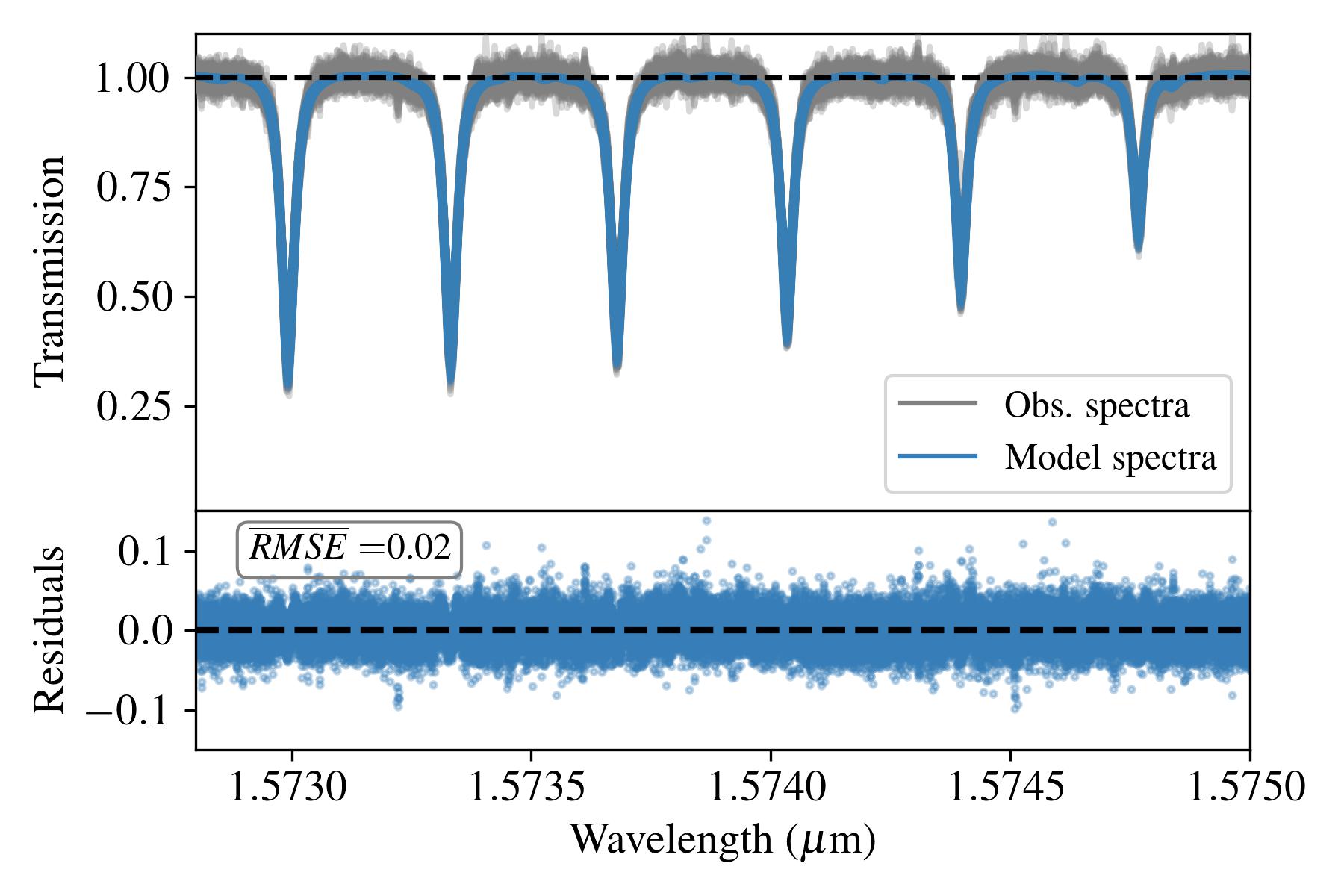}
    \caption{Top: best-fit model telluric transmission spectra computed with \textit{Astroclimes} (blue) and observational spectra from CARMENES (grey) for a portion of the CARMENES order 25, which contains \ce{CO2} lines. Bottom: residuals plot (observed spectra minus model spectra). The RMSE value shown on the top left corner is the mean RMSE for all observations, and it should be noted that each RMSE is calculated for the whole spectral range used, (i.e. all CARMENES orders used), not just the range shown in the figure.}
    \label{fig_ex_spec}
\end{figure}

An example of how the model spectra look compared to the observational spectra is shown for a wavelength region containing \ce{CO2} lines in Figure \ref{fig_ex_spec}. The match between the observational and model spectra is quantified by the root-mean-squared error (RMSE). The value shown at the bottom panel of Figure \ref{fig_ex_spec}, $\overline{\text{RMSE}}=0.02$, is the mean of all of the RMSE for each observation and indicates that overall there has been a good agreement between the model spectra and their observational counterpart.

On average, the uncertainties on our individual measurements fell within $0.3-0.7\%$, that is $1-2$ ppm for \ce{CO2}, $11-13$ ppb for \ce{CH4}, $6-18$ ppm for \ce{H2O} and around 650 ppm for \ce{O2}. However, these are simply precision errors based on the 68\% confidence intervals of the MCMC analysis posterior distribution, they do not include any systematic biases nor any scaling related to known error sources. Identifying and correcting systematic biases is a lengthy and ongoing process that is required even in large measurement networks and collaborations, such as for the OCO-2/OCO-3 \citep{odell2018} and GOSAT-2 \citep{yoshida2023} satellites. Even TCCON, which is used to calibrate the aforementioned satellites \citep{wunch2011b}, requires a calibration of its own to be in agreement with the World Meteorological Organization Global Atmosphere Watch (WMO/GAW) \ce{CO2} calibration scale \citep{hall21}, a process that is done via aircraft measurements over TCCON sites \citep{wunch2010}. Naturally, it is expected that the retrievals reported here have biases as well, and a systematic offset and precision scatter were quantified in this work, but further biases and improvements will be investigated in future work.

For context, according to \cite{miller2007}, a precision of 1-2 ppm is needed for $X_\text{\ce{CO2}}$ measurements ``on regional scales to improve our knowledge of carbon cycle phenomena'', but even the largest sinks and sources of \ce{CO2} rarely produce changes in the background $X_\text{\ce{CO2}}$ distribution that exceed 0.25\% \citep{crisp17}. On the other hand, \cite{meirink2006} concluded that $X_\text{\ce{CH4}}$ measurements with a precision of 1 to 2\% can contribute considerably to reduce the uncertainty in the strength of methane sources, and systematic errors below 1\% have a dramatic impact on the quality of the derived emission fields. 

\begin{figure}
    \centering
    \includegraphics[width=\columnwidth]{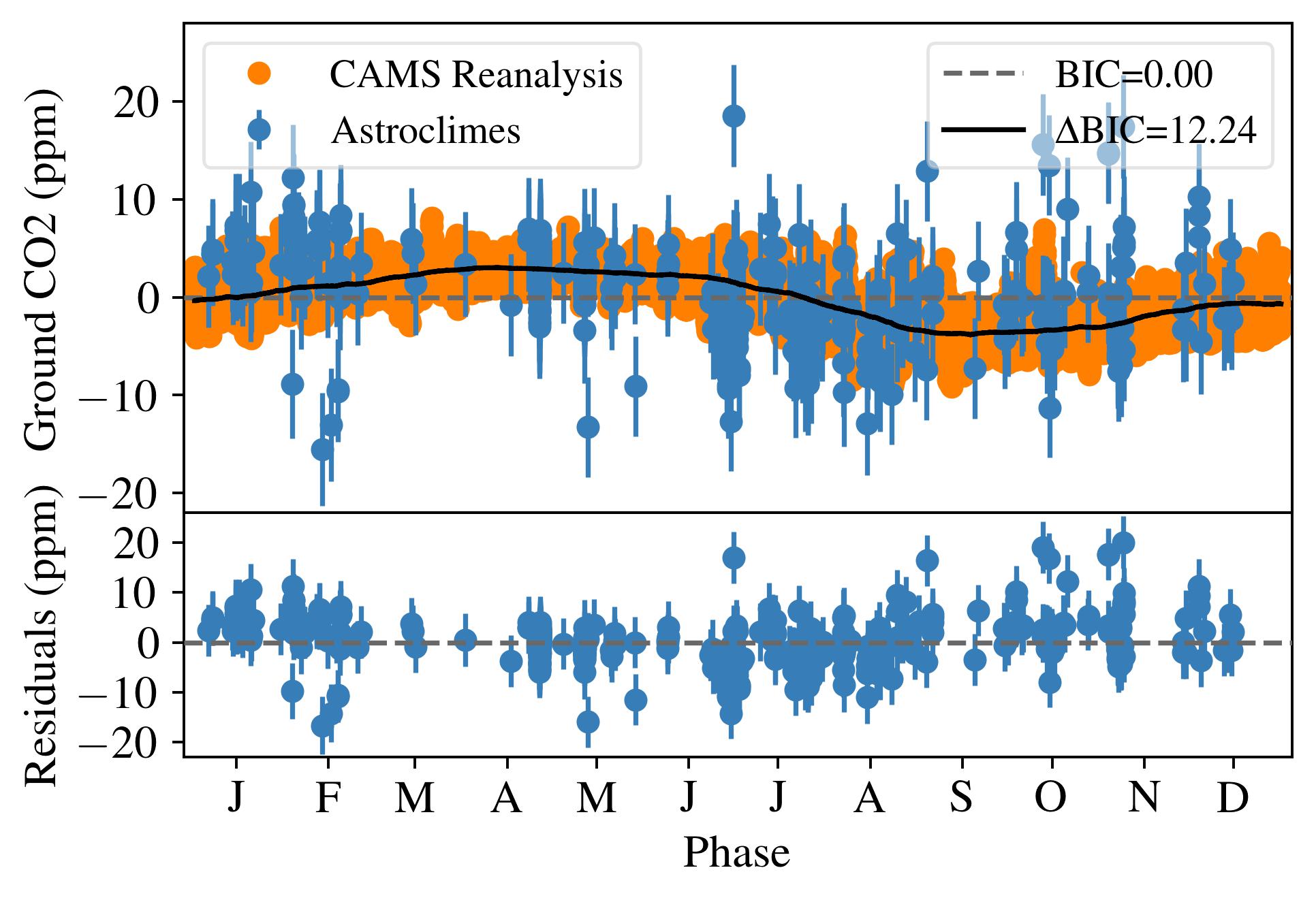}
    \caption{Top: phase-folded ground level \ce{CO2} DMFs retrieved with \textit{Astroclimes} (blue) and from the CAMS global greenhouse gas reanalysis model (orange). The solid black line is a rolling median of the CAMS values with a window size of 10\% the number of points. The BIC of the Astroclimes retrievals compared to the solid black line is reported on the top right and is compared to the BIC of the Astroclimes retrievals and a horizontal line centred on 0 (dashed gray line). The values shown as the $y$-axis correspond to the seasonal variation of the abundances with the long-term trend removed based on a fitted function as described in the text. Bottom: residuals between the retrieved \textit{Astroclimes} values and the interpolated rolling median of the CAMS values. The $x$-axis tick labels correspond to the months, with each tick placed at the 15th of each month.}
    \label{fig_CO2_phase_fold}
\end{figure}

\begin{figure}
    \centering
    \includegraphics[width=\columnwidth]{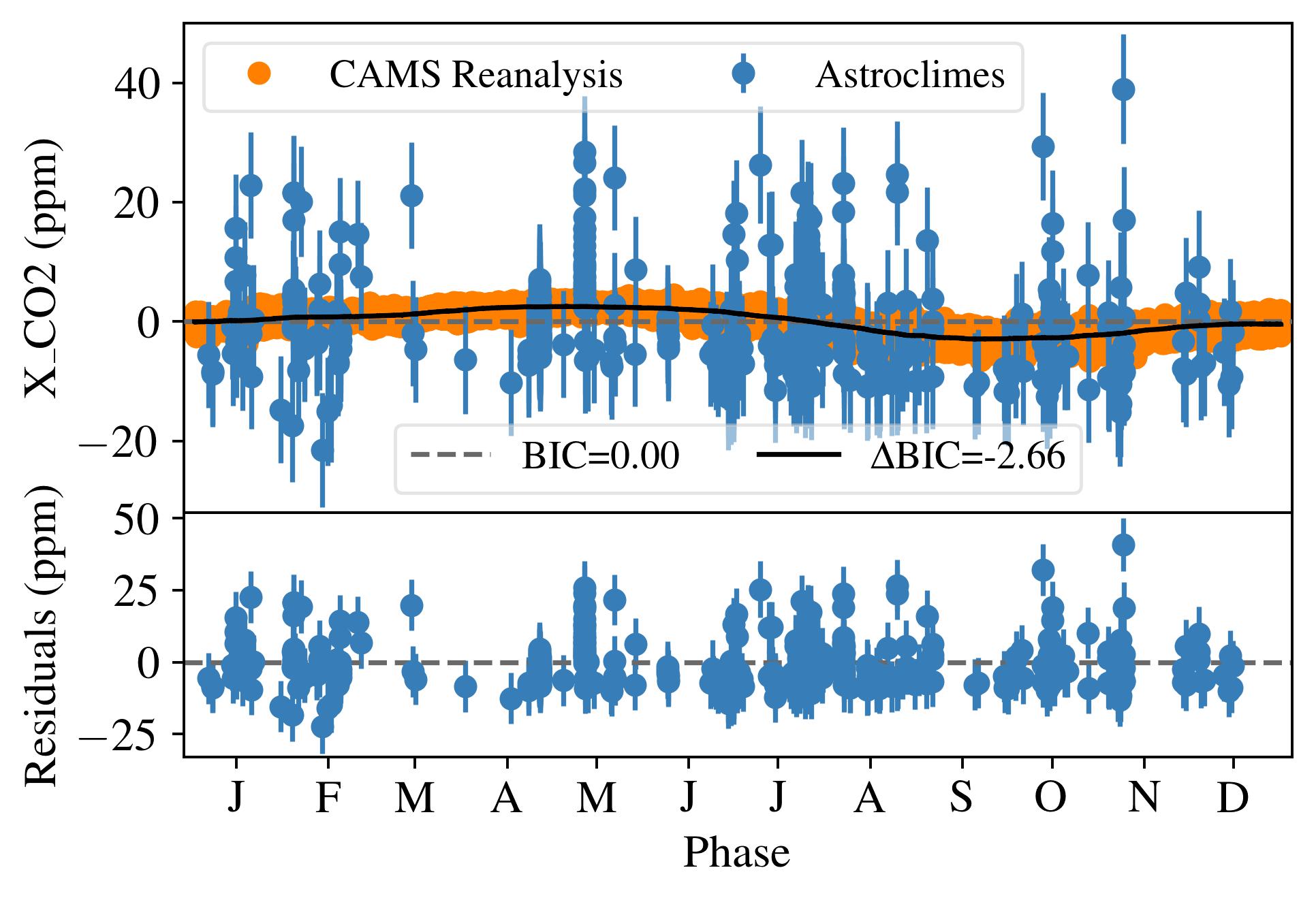}
    \caption{Same as Figure \ref{fig_CO2_phase_fold}, but for the column-averaged \ce{CO2} DMFs instead.}
    \label{fig_XCO2_phase_fold}
\end{figure}

\begin{figure}
    \centering
    \includegraphics[width=\columnwidth]{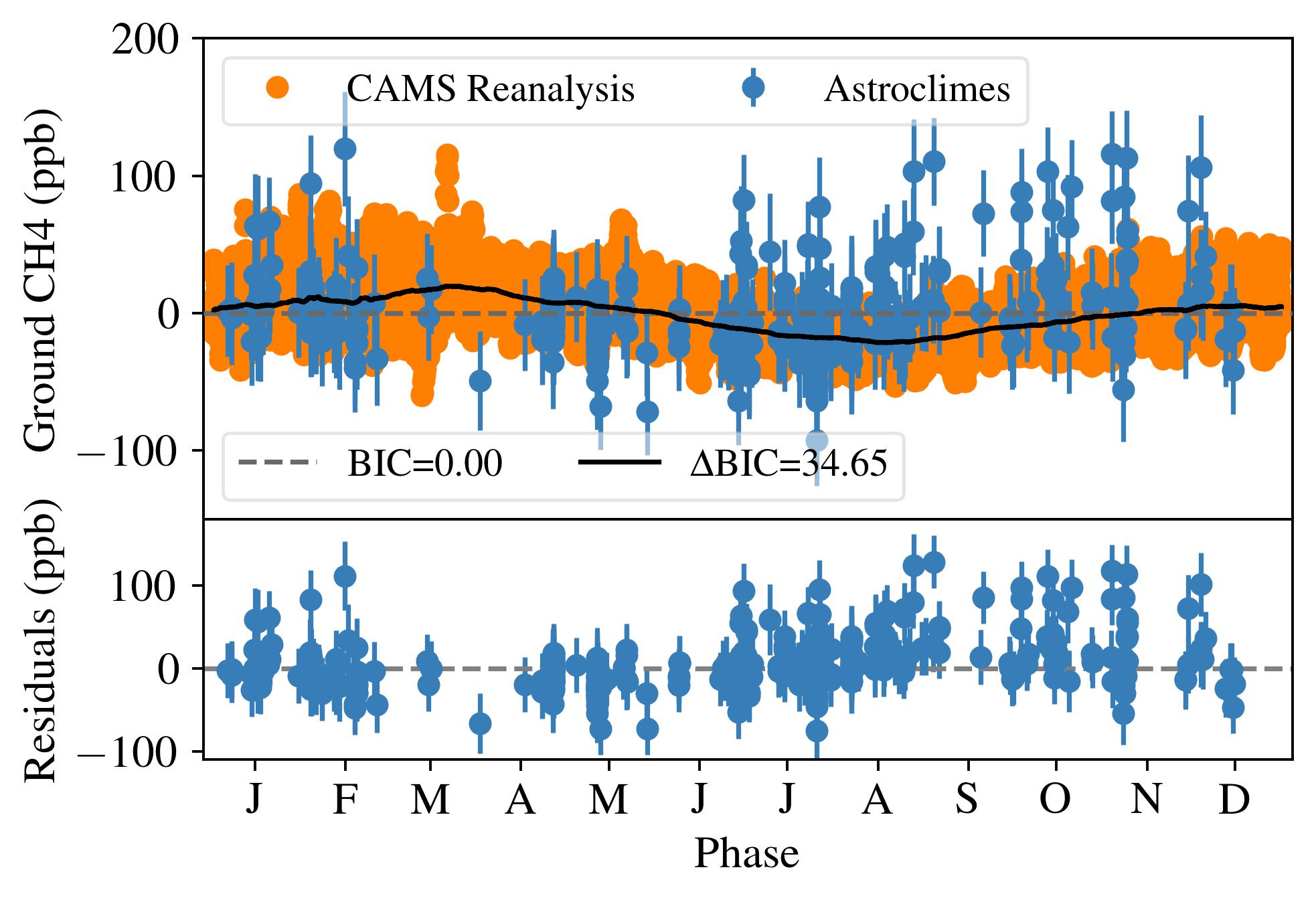}
    \caption{Same as Figure \ref{fig_CO2_phase_fold}, but for the ground level \ce{CH4} DMFs instead.}
    \label{fig_CH4_phase_fold}
\end{figure}

\begin{figure}
    \centering
    \includegraphics[width=\columnwidth]{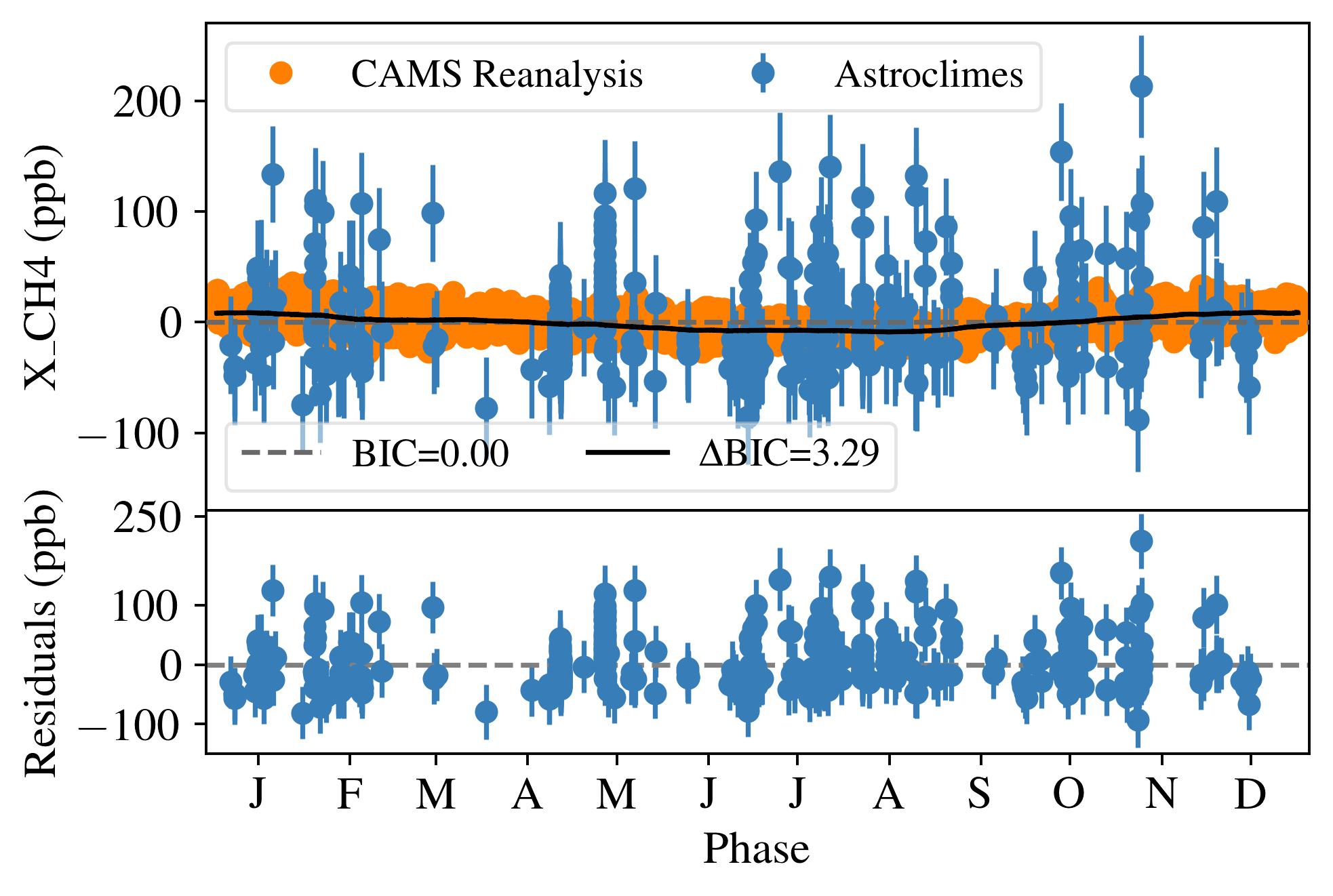}
    \caption{Same as Figure \ref{fig_CO2_phase_fold}, but for the column-averaged \ce{CH4} DMFs instead.}
    \label{fig_XCH4_phase_fold}
\end{figure}

\begin{figure*}
    \centering
    \includegraphics[width=\textwidth]{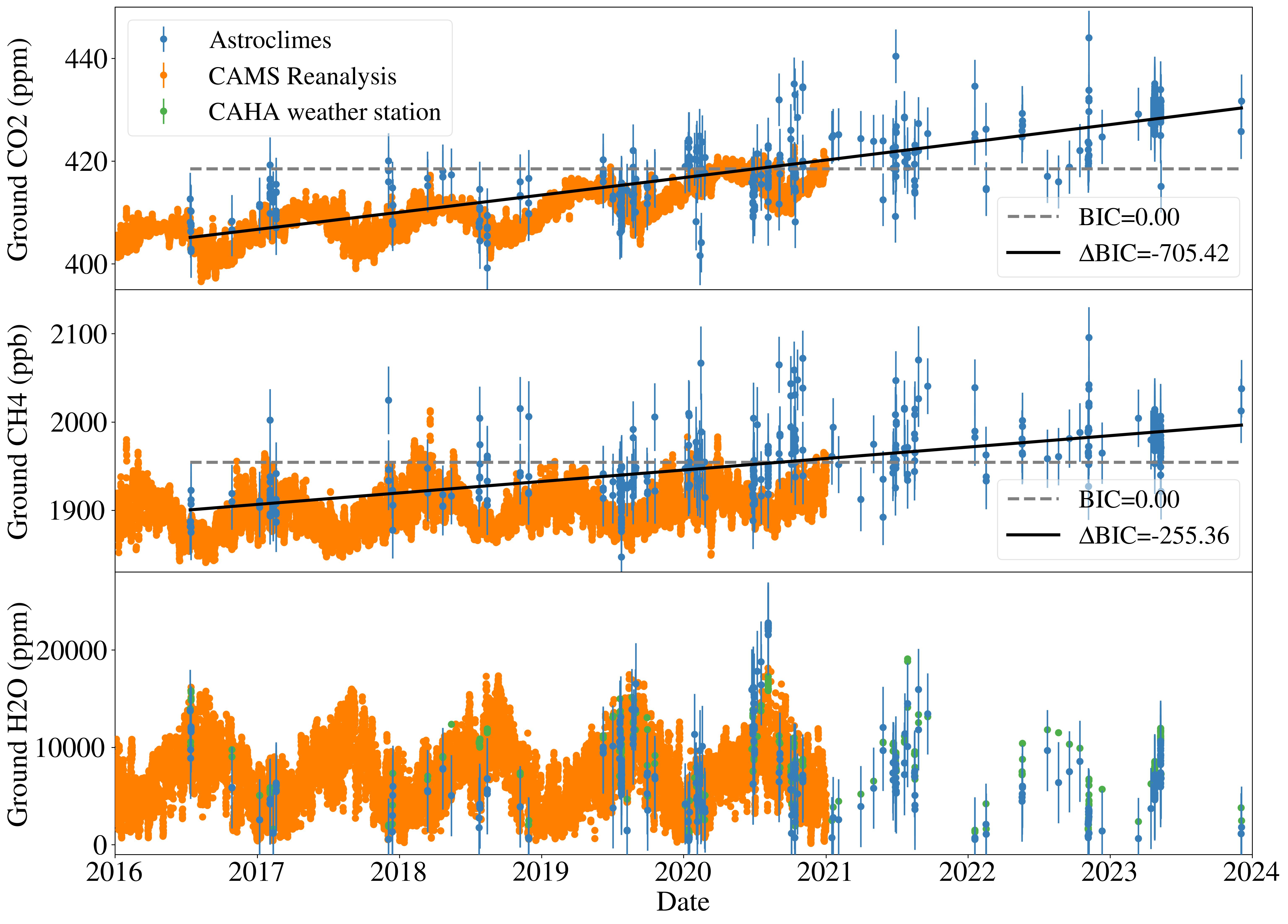}
    \caption{Ground level DMFs retrieved by \textit{Astroclimes} (blue) and from the CAMS global greenhouse gas reanalysis model (orange) for \ce{CO2} (top), \ce{CH4} (middle) and \ce{H2O} (bottom), the latter which also shows the humidity measurements taken by the CAHA weather station (green), converted to ppm based on the measured temperature and pressure at the time of observation using equation (\ref{eq_pwv}). For \ce{CO2} and \ce{CH4}, the \textit{Astroclimes} values have been vertically shifted to match the CAMS data based on the $c_0$ values obtained in the long-term trend fitting function described in the text. The BICs of the Astroclimes \ce{CO2} and \ce{CH4} retrievals are reported with respect to the same fits used to remove the long term trend (solid black lines) and is compared to the BICs of the Astroclimes retrievals and a horizontal line centred on their mean value (dashed gray lines).}
    \label{fig_ground_evols}
\end{figure*}

\begin{figure*}
    \centering
    \includegraphics[width=\textwidth]{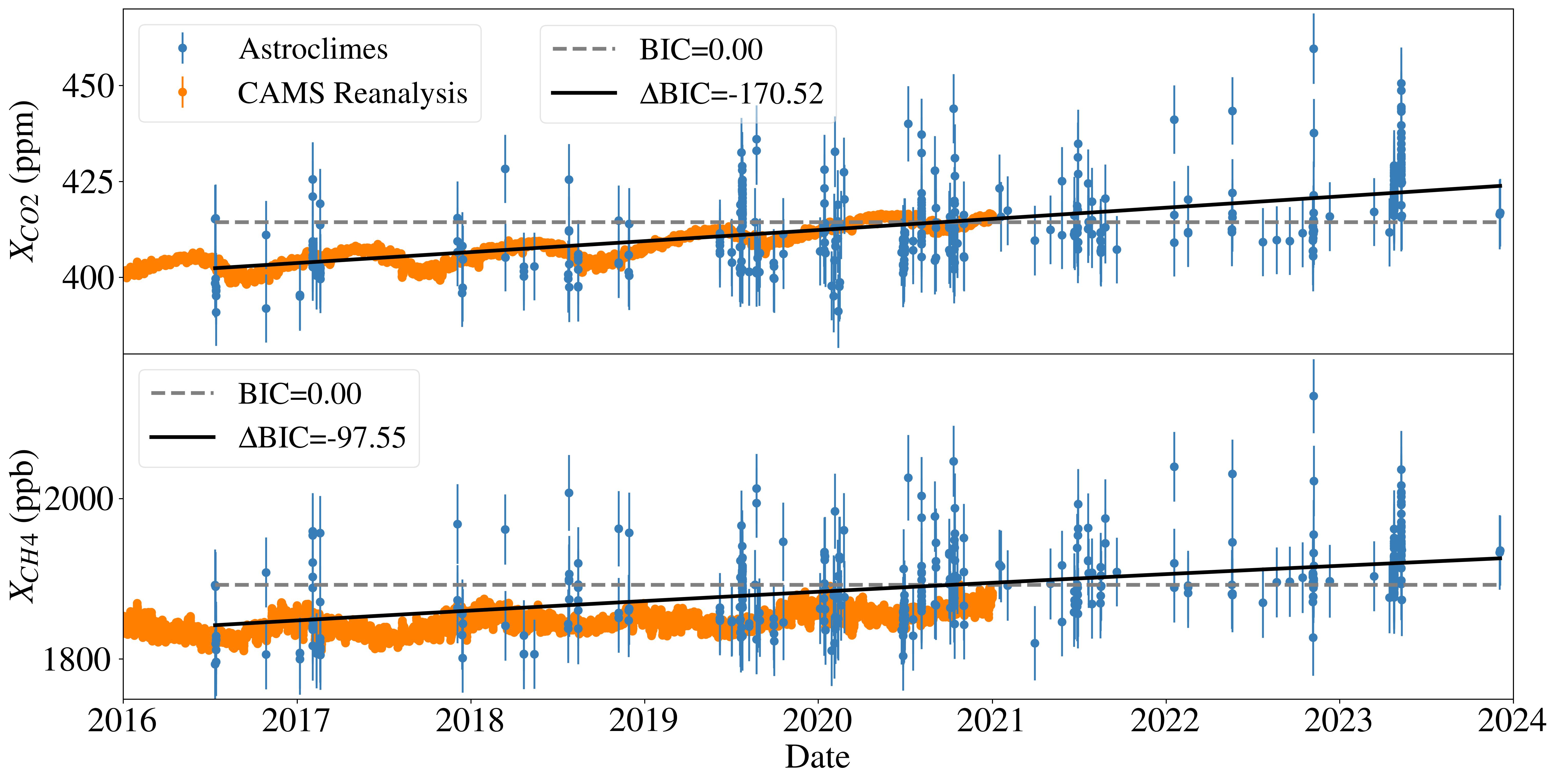}
    \caption{Column-averaged DMFs retrieved by \textit{Astroclimes} (blue) and from the CAMS global greenhouse gas reanalysis model (orange) for \ce{CO2} (top) and \ce{CH4} (bottom) and \ce{H2O} (bottom). The \textit{Astroclimes} values have been vertically shifted to match the CAMS data based on the $c_0$ values obtained in the long-term trend fitting function described in the text. The BICs reported have the same meaning as in Figure \ref{fig_ground_evols}.}
    \label{fig_X_evols}
\end{figure*}

If the uncertainties for the individual measurements were to be trusted, they would in theory fall within the precision requirements to detect sinks and sources of \ce{CO2} and \ce{CH4}. However, as evidenced by Figures \ref{fig_CO2_phase_fold}-\ref{fig_X_evols}, the large spread shown in the results indicates that these uncertainties are at this point underestimated. Therefore, the CAMS global greenhouse gas reanalysis EGG4 was used as a benchmark to quantify the scatter in our results and provide a more realistic uncertainty for our retrievals. 

This process was carried out by first fitting a function to remove any long-term trend present, then phase-folding the data with a period of 365 days and calculating a rolling median to highlight the seasonal trend. The fitting functions were chosen to follow the same approach as for the GGG2020 pipeline when extrapolating their atmospheric profiles, so equations (\ref{eq_CO2_fit_func}) and (\ref{eq_CH4_fit_func}) were used for \ce{CO2} and for \ce{CH4}, respectively. A small change was applied as instead of fitting for $t$, we fitted for $t-T_0$, where $T_0 = 2457388.5$ BJD, or midnight on 01/01/2016, which is the start date we chose for the CAMS reanalysis data. This way, $c_0$ would correspond to the function value on the first day of 2016.

The phase-folded CAMS distribution was then interpolated to the time grid of the CARMENES observations so they could be directly compared. The scatter on the retrievals was determined to be the standard deviation of the residuals between the retrieved values and the interpolated CAMS values. Figures \ref{fig_CO2_phase_fold}-\ref{fig_XCH4_phase_fold} show the results of this analysis, where the uncertainties in the Astroclimes retrievals are obtained by adding in quadrature the individual uncertainties from the MCMC and the scatter when compared to the CAMS values. 

For the ground level \ce{CO2} abundances, the scatter analysis yielded a value of $\pm 5$ ppm for the standard deviation of the residual distribution, while for the column-averaged DMFs this value was $\pm 9$ ppm. For the ground level and column-averaged \ce{CH4} DMFs, they were $\pm 31$ ppb and $\pm 42$ ppb. The larger scatter exhibited by the column-averaged DMFs could be due to the scatter in the retrieved column-averaged \ce{O2} DMFs, which would be expected to be constant. We hypothesise that the reason for this might be due to the fact that in the spectral region covered by the CARMENES NIR range only contains one set of \ce{O2} lines (at around $1.27\ \mu$m), which are quite weak, so they may compromise our ability to properly retrieve \ce{O2} abundances. The CARMENES visible range, however, contains the stronger \ce{O2} A-band (at around $0.76\ \mu$m), so a potential future improvement would be to combine this dataset to check if there are any improvements in the retrievals.

By fitting functions to the long term trend of \ce{CO2} and \ce{CH4}, it became evident that our retrieved values were consistently overestimated compared to the CAMS reanalysis model. The reason for this overall discrepancy is currently unknown and will be investigated for future work. Therefore, when comparing our results with the CAMS reanalysis data, we apply a shift to our retrieved values based on $c_0$ values obtained in the long-term trend fitting function. For the \ce{CO2} ground level and column-averaged DMFs, the shifts were 14 and 15 ppm, respectively, while for \ce{CH4} they were 42 and 7 ppb, respectively. Figures \ref{fig_ground_evols} and \ref{fig_X_evols} show the \textit{Astroclimes} retrievals for the full CARMENES dataset compared to the CAMS reanalysis model.

We used the Bayesian information criterion (BIC) to assess the match between the Astroclimes retrievals and the long-term trend and seasonal cycle exhibited by \ce{CO2} and \ce{CH4}. For the distributions of both molecules, the long term trend is present in the \textit{Astroclimes} retrievals, as evidenced by the lower BIC of the solid black lines on Figures \ref{fig_ground_evols} and \ref{fig_X_evols}, which is promising, despite the larger scatter and overall vertical shift when compared to the values from the CAMS reanalysis model. We found that the change in BIC is not significant enough to claim to have recovered the seasonal cycle of either \ce{CO2} or \ce{CH4} (Figures \ref{fig_CO2_phase_fold}-\ref{fig_XCH4_phase_fold}). In order to reduce the scatter and properly correct the vertical shift, more work is needed to identify and mitigate potential systematic biases and error sources. 

Unlike \ce{CO2} and \ce{CH4}, \ce{H2O} is known and expected to have a variable abundance in very short timescales (minutes to hours). This is exemplified by the much larger scatter in the CAMS reanalysis points from the bottom panel of Figure \ref{fig_ground_evols} compared to the scatter on their values for \ce{CO2} and \ce{CH4}. This panel also contains the humidity measurements from the CAHA weather station, located close to the telescope that hosts the CARMENES spectrograph, which provides a more direct comparison for our results, even though we still rely on model-assisted atmospheric profiles to calculate the ground level abundances. The \textit{Astroclimes} retrieved values show reasonable agreement with both of these distributions, thus attesting the algorithm's capability to track rapid changes in molecular abundances.

\subsubsection{Results from individual nights}
Despite the uncertainties on the individual \textit{Astroclimes} retrievals ranging from 0.3-0.7\% for \ce{CO2} and \ce{CH4}, the scatter on the whole sample seems to be larger (1-2\%) when compared to the CAMS reanalysis values. We also investigated the scatter on shorter timescales for observations within the same night. On the whole CARMENES sample, there are two nights where the same target was observed for an extended period: one is July 24th 2019, which has 37 observations of MASCARA-1; the other is April 24th 2023, which has 73 observations of HR 5676 (also known as the weather balloon night). 

\begin{figure}
    \centering
    \includegraphics[width=\columnwidth]{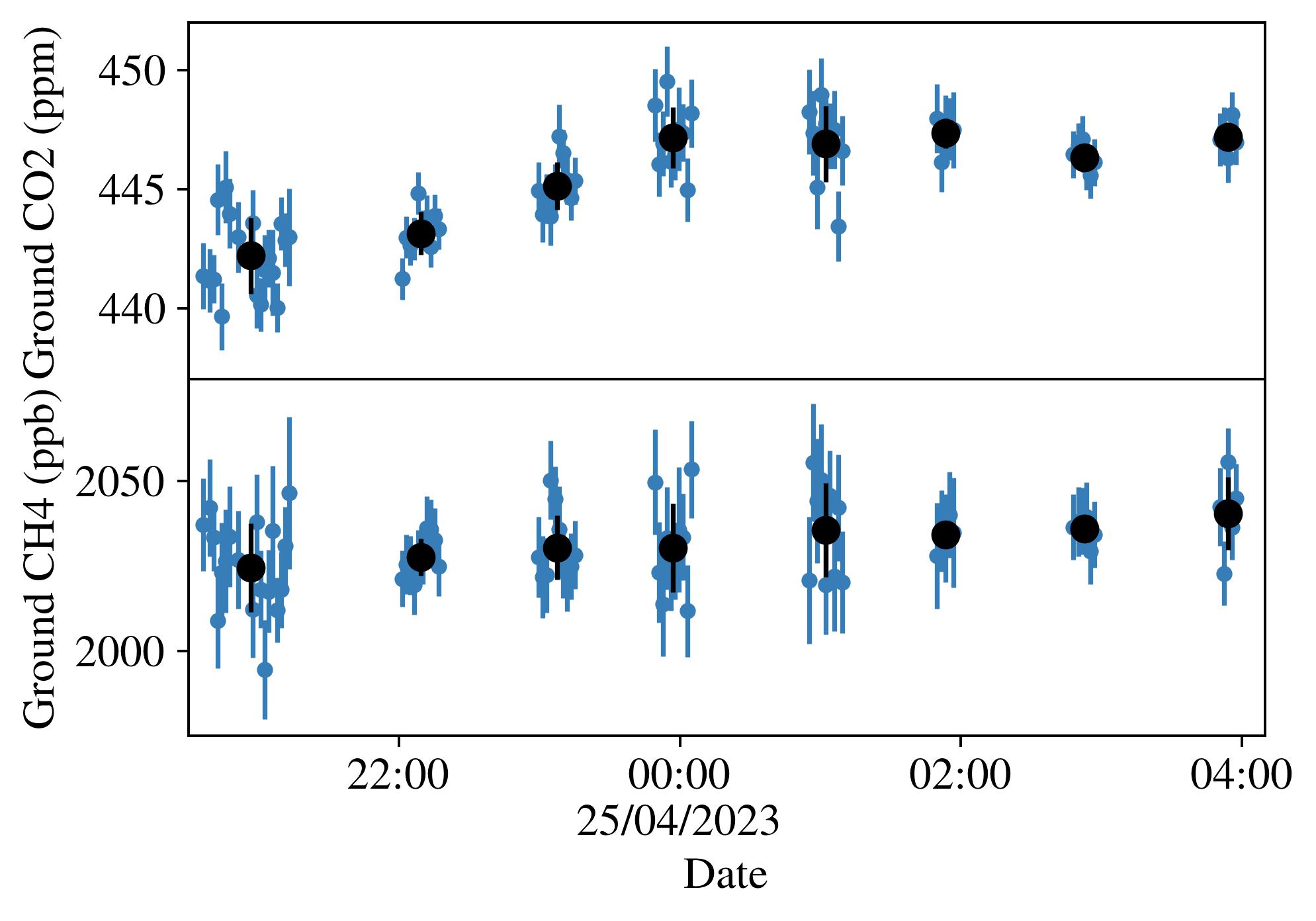}
    \caption{Ground level DMFs retrieved by \textit{Astroclimes} (blue) for the observations of HR5676 on the night of April 24th 2023. The black points correspond to the mean of the retrievals binned by hour, whose uncertainties are the standard deviation of the retrievals in said bin. The uncertainties on the individual points are those from the MCMC posterior distribution.}
    \label{fig_ground_evols_2023_04_24}
\end{figure}

\begin{figure}
    \centering
    \includegraphics[width=\columnwidth]{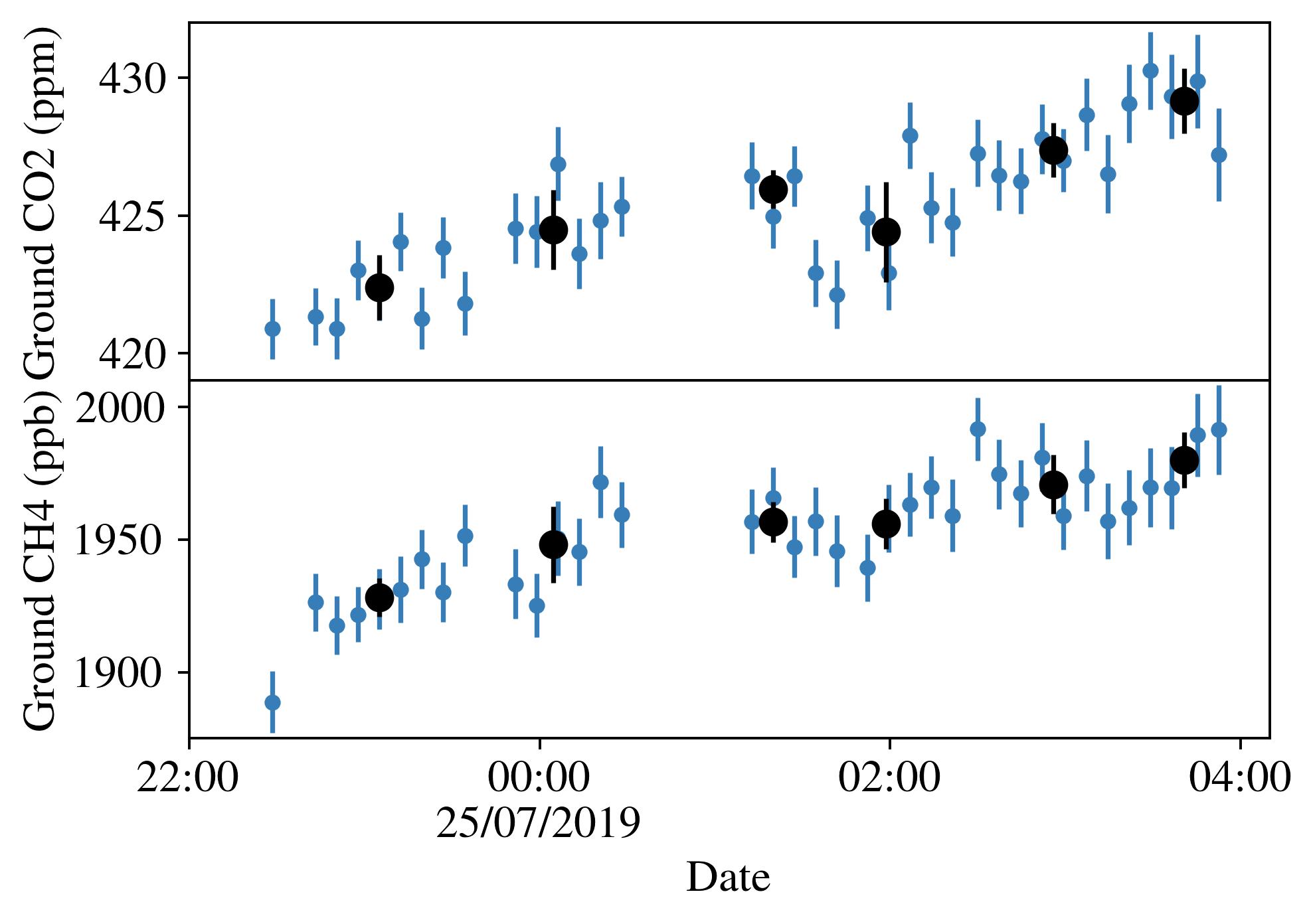}
    \caption{Same as Figure \ref{fig_ground_evols_2023_04_24}, but for the observations of MASCARA-1 on the night of July 24th 2019.}
    \label{fig_ground_evols_2019_07_24}
\end{figure}

These two nights were used to assess the variability in the retrieval of \ce{CO2} and \ce{CH4}. The observations of each night were binned in hourly clumps, and the mean of the retrievals in each clump was calculated, as well as the standard deviation. The ground level \ce{CO2} and \ce{CH4} retrievals, along with the aforementioned binned values, are shown in Figures \ref{fig_ground_evols_2023_04_24} and \ref{fig_ground_evols_2019_07_24} for the nights of April 24th 2023 and July 24th 2019, respectively.

For ground \ce{CO2}, the mean of the standard deviation of the hourly clumps in each night was $\pm 1$ ppm and $\pm 1.2$ ppm for the HR 5676 night and the MASCARA-1 night, respectively, while for ground \ce{CH4} these values were $\pm 9$ ppb and $\pm 10$ ppb, respectively, which are on par with the individual uncertainties of the \textit{Astroclimes} retrievals. 

The fact that the scatter on short timescales is in line with the MCMC precision errors implies that the additional scatter compared to the CAMS baseline is due to longer term systematic errors and/or physical variation. If this longer term variation can be understood, the data quality and method has the potential to reach state-of-the-art precision.

It is also clear that both nights exhibit a trend in the retrievals that cannot be attributed to the noise level, but at this point it cannot be confirmed whether this trend is physical in origin or if it is due to uncorrected systematic biases.

\section{Conclusions}\label{conclusions}

We present a new method to measure the column-averaged DMF of greenhouse gases in the Earth's atmosphere using our newly developed synthetic transmission modelling and fitting algorithm named \textit{Astroclimes}. While most current ground and space-based networks only provide column measurements of \ce{CO2} and \ce{CH4} during the day, as they rely on sunlight, with \textit{Astroclimes} we can carry out nighttime measurements using telluric standard stars. Besides providing column measurements of these gases at night, we can also provide measurements in places that may not be covered by the existing networks. 

Any type of ground-based spectroscopic astronomical observation in the near-infrared can in principle be used to retrieve the column-averaged dry air mole fraction of \ce{CO2} and \ce{CH4}. Relevant data is available in archives and continuously being taken at multiple observatories.

To validate our algorithm, we tested it against spectra computed with the ESO Sky Model Calculator, SKYCALC. We found that \textit{Astroclimes} was able to reproduce the reported PWV values up to 20 mm.

An MCMC analysis was performed on an extensive dataset from the CARMENES spectrograph. \textit{Astroclimes} was able to reproduce the long-term trend in both the \ce{CO2} and \ce{CH4} distributions, but not their seasonal cycles. Benchmarking against the CAMS global greenhouse gas reanalysis model EGG4 quantified the scatter in our retrievals for the ground level and column-averaged \ce{CO2} DMFs to be $\pm 5$ ppm and $\pm 9$ ppm, respectively. For \ce{CH4}, these values were $\pm 31$ ppb and $\pm 42$ ppb, respectively. In addition, our retrievals also exhibited an overall vertical shift when compared to the CAMS reanalysis values. This shift was around 14 ppm and 15 ppm for the \ce{CO2} ground level and column-averaged DMFs, respectively, and 42 ppb and 7 ppb for the \ce{CH4} ground level and column-averaged DMFs, respectively. 

Two nights that contained numerous observations of a single target were selected to quantify the nightly scatter in the \textit{Astroclimes} retrievals. These proved to be lower than the long-term scatter determined by the comparison with the CAMS values. For the ground \ce{CO2} retrievals, the nightly scatter was of order $\pm 1$ ppm, while for ground \ce{CH4} it was of order $\pm 10$ ppb, which is on par with the uncertainties on individual measurements.

These values are of lower precision than current state-of-the-art, which can get to sub-ppm precision, but demonstrate the feasibility of recovering information on greenhouse gas abundances with astronomical observations. Further work remains to be done to improve calibration and increase precision.

\section*{Acknowledgements}

This research was funded in part by the UKRI (Grants ST/X001121/1, EP/X027562/1, MR/S035214/1) and by the Astronomy and Astrophysics Warwick Prize Scholarship. 

We thank the Calar Alto Observatory for allocation of director's discretionary time to this programme (DDT.23A.304) and the CAHA staff for carrying out the astronomical observations and showing flexibility and excitement towards our weather balloon experiment, as well as the people from the ``Comunidad de Regantes de Águilas'' for retrieving our weather balloon and getting in touch with us. 

MAFK would also like to thank Thorsten Warneke and Christof Petri from the University of Bremen and PIs of the TCCON Orléans and Nicosia sites, respectively, for their guidance and assistance on generating the GGG2020 atmospheric profiles and being open to future collaborations.

\section*{Data Availability}
This article uses or references data from many sources. The molecular cross-sections tables and the weather balloon data were computed/measured by the authors and are not publicly available but can be shared upon reasonable request to the corresponding author. The CARMENES data can be obtained from the CAHA Archive at \href{http://caha.sdc.cab.inta-csic.es/calto/}{http://caha.sdc.cab.inta-csic.es/calto/}. The SKYCALC model spectra can be generated at \href{https://www.eso.org/sci/software/pipelines/skytools/skymodel}{https://www.eso.org/sci/software/pipelines/skytools/skymodel}. The MIPAS atmospheric profiles can be downloaded from \href{https://eodg.atm.ox.ac.uk/RFM/atm/}{https://eodg.atm.ox.ac.uk/RFM/atm/}. The GDAS atmospheric profiles are provided by the NOAA and can be downloaded from the \textit{Molecfit} local database at \href{https://ftp.eso.org/pub/dfs/pipelines/skytools/molecfit/gdas/}{https://ftp.eso.org/pub/dfs/pipelines/skytools/molecfit/gdas/}. The GGG2020 atmospheric profiles were computed by the authors using \texttt{ginput} and can be shared upon reasonable request to the corresponding author. The GEOS data used in this paper have been provided by the Global Modelling and Assimilation Office (GMAO) at NASA Goddard Space Flight Center through the online data portal in the NASA Center for Climate Simulation at \href{https://portal.nccs.nasa.gov/datashare/gmao/geos-fp/das/}{https://portal.nccs.nasa.gov/datashare/gmao/geos-fp/das/}. The Copernicus Atmosphere Monitoring Service (CAMS) global reanalysis data used in this paper was generated and downloaded from the CAMS Atmosphere Data Store (ADS) at \href{https://ads.atmosphere.copernicus.eu/}{https://ads.atmosphere.copernicus.eu/}. 



\bibliographystyle{mnras}
\bibliography{refs} 




\appendix

\begin{landscape}
\section{Rules included in the normalisation and data handling process}\label{appendix_A}

\begin{table}
    \centering
    \begin{tabular}{|c|c|c|c|}
        \hline 
        Rule number & What is the rule & Parameters & Values \\
        \hline
         1 & Defines line and continuum points & Limiting value for continuum points & > -1$\times$ MAD \\
         \hline
         2 & Molecular abundances for dummy model & Ground level \ce{CO2} and \ce{H2O} abundances & $8\times 10^{21} \text{molecules}/\text{m}^3$, $1.5\times 10^{23} \text{molecules}/\text{m}^3$\\
         \hline
         3 & 1st median filter window sizes & Fraction of total number of points & [0.10, 0.15, 0.10, 0.10, 0.10, 0.10] \\
        \hline
         4 & 2nd median filter window sizes & Fraction of total number of points & [0.03, 0.01, 0.01, 0.01, 0.01, 0.01] \\
         \hline
         5 & Defines emission line positions & Minimum emission line height & [500, 100000, 40000, 4000, 2000, 3000]*, in photons~s$^{-1}$~m$^{-2}$~micron$^{-1}$arcsec$^{-2}$ \\
         \hline
         6 & Defines minimum transmission included in fit & Limiting normalised flux for absorption lines & > 0.2 \\
         \hline
    \end{tabular}
    \caption{Summary of the parameters used for the rules applied in the normalisation and data handling process. The parameters in square brackets represent the values for each of the CARMENES orders used. \\
    * These values were determined based on an emission spectra computed for an observatory height of 2640m (Cerro Paranal), airmass of 1, season and period of night are entire year and entire night, PWV = 2.5mm, a resolution of 20000 and wavelength range from $1.0\mu m - 1.8\mu m$.}
    \label{tab_algorithm_rules}
\end{table}
\end{landscape}


\bsp	
\label{lastpage}
\end{document}